\documentclass[twocolumn,showpacs,preprintnumbers,amsmath,amssymb,superscriptaddress,nofootinbib,english]{revtex4-1}

\usepackage{graphicx}
\usepackage{dcolumn}
\usepackage{bm}
\usepackage{epsfig}
\usepackage{graphicx}
\usepackage{hyperref}
\usepackage[usenames]{color}
\usepackage{url}

\hypersetup{
    colorlinks=true,
    linkcolor=red,
    citecolor=blue,
}

\newcommand{\m}{{\rm M}}

\newcommand{\ab}{\alpha\beta}
\newcommand{\fR}{f_{R}}
\newcommand{\dfR}{{\delta}f_{R}}

\newcommand{\rd}{{\rm d}}
\newcommand{\dr}{\delta{R}}
\newcommand{\remove}[1]{}

\def\etal{{\frenchspacing\it et al.}}
\def\ie{{\frenchspacing\it i.e.}}
\def\eg{{\frenchspacing\it e.g.}}
\def\etc{{\frenchspacing\it etc.}}

\def\be{\begin{equation}}
\def\ee{\end{equation}}
\def\ba{\begin{eqnarray}}
\def\ea{\end{eqnarray}}

\begin{document}

\title{$N$-body Simulations for $f(R)$ Gravity using a Self-adaptive Particle-Mesh Code}

\author{Gong-Bo~Zhao}
\email[Email address: ]{Gong-Bo.Zhao@port.ac.uk}
\affiliation{Institute of Cosmology \& Gravitation, University of Portsmouth, Dennis Sciama Building, Portsmouth, PO1 3FX, UK}

\author{Baojiu~Li}
\email[Email address: ]{B.Li@damtp.cam.ac.uk}
\affiliation{DAMTP, Centre for Mathematical Sciences, University of Cambridge,
Wilberforce Road, Cambridge CB3 0WA, UK}
\affiliation{Kavli Institute for Cosmology Cambridge, Madingley Road, Cambridge CB3 0HA,
UK}
\author{Kazuya~Koyama}
\email[Email address: ]{Kazuya.Koyama@port.ac.uk}
\affiliation{Institute of Cosmology \& Gravitation, University of Portsmouth, Dennis Sciama Building, Portsmouth, PO1 3FX, UK}

\date{\today}

\begin{abstract}

We perform high resolution N-body simulations for $f(R)$ gravity based on a self-adaptive particle-mesh code {\tt MLAPM}.
The Chameleon mechanism that recovers General Relativity on small scales is fully taken into account by self-consistently solving the non-linear equation for the scalar field. We independently confirm the previous simulation results, including the matter power spectrum, halo mass function and density profiles, obtained by Oyaizu \etal~(Phys.Rev.D {\bf 78}, 123524, 2008) and Schmidt \etal~(Phys.Rev.D {\bf 79}, 083518, 2009), and extend the resolution up to $k\sim20$ h/Mpc for the measurement of the matter power spectrum. Based on our simulation results, we discuss how the Chameleon mechanism affects the clustering of dark
matter and halos on full non-linear scales.

\end{abstract}

\maketitle

\section{Introduction}

The late-time acceleration of the Universe is the most
challenging problem in cosmology. Within the framework of general
relativity (GR), the acceleration originates from dark energy with negative pressure.
The simplest candidate for dark energy is the cosmological constant. However, in order to
explain the current acceleration of the Universe, the required value of
the cosmological constant must be incredibly small. Alternatively, there
could be no dark energy, but a large scale modification of GR may
account for the late-time acceleration of the Universe. Recently considerable
efforts have been made to construct models for modified gravity as an
alternative
to dark energy, and to distinguish them from dark energy models by observations
(see \cite{Nojiri:2006ri, Koyama:2007rx, Durrer:2008in, Durrer:2007re, Jain:2010ka} for
reviews).

Although fully consistent models have not been constructed yet,
some indications of the nature of the modified gravity models
have been obtained. In general, there are three regimes of gravity
in modified gravity models \cite{Koyama:2007rx, Hu07}.
On the largest scales, gravity must be modified
significantly in order to explain the late time acceleration without
introducing dark energy. On the smallest scales, the theory must approach GR
because there exist stringent constraints on the deviation from GR at solar
system scales. On intermediate scales between the cosmological horizon
scales and the solar system scales, there can be still a deviation
from GR. In fact, it is a very common feature in modified gravity models
that gravity can deviate from GR significantly on large scales. This is due to the fact that, once we modify GR, there arises a new
scalar degree of freedom in gravity. This scalar mode changes gravity even
below the length scale where the modification of gravity becomes significant
that causes the cosmic acceleration.

In order to recover GR on small scales, it is required to suppress
the interaction arising from this scalar mode. One of the well known mechanisms
is the so-called Chameleon mechanism \cite{Khoury:2003rn}. A typical example is $f(R)$ gravity where
the Einstein-Hilbert action is replaced by
an arbitrary function of Ricci curvature (see \cite{Sotiriou:2008rp, DeFelice:2010aj}
for a review). This model is equivalent to the Brans-Dicke (BD) theory with a non-trivial
potential. The BD scalar mediates an additional gravitational interaction, which enhances the gravitational force below the Compton wavelength of the BD scalar. If the mass of the BD scalar becomes larger in a dense
environment like in the solar system, the Compton wavelength becomes
short and GR can be recovered. This is known as the Chameleon mechanism
\cite{Khoury:2003rn}. In the context of $f(R)$ gravity, by tuning the function $f$,
it is possible to make the Compton wavelength of the BD scalar short at solar system
scales and screen the BD scalar interaction
\cite{Hu:2007nk, Starobinsky:2007hu, Appleby:2007vb, Cognola:2007zu}.

This mechanism affects the non-linear clustering of dark matter. We naively
expect that the power-spectrum of dark matter perturbations approaches the one in the $\Lambda$CDM model with the same expansion history of the Universe because the modification
of gravity disappears on small scales. Then the difference between a modified gravity model
and a dark energy model with the same expansion history becomes smaller on smaller scales.
This recovery of GR has important implications for weak lensing measurements
because the strongest signals in weak lensing measurements come from
non-linear scales. One must carefully take into account this effect when constraining
the model in order not to over-estimate the deviation from GR.

Using a concrete example in $f(R)$ gravity, Refs.~\cite{Oyaizu:2008sr, Oyaizu:2008tb, Schmidt:2008tn}
performed N-body simulations to confirm this expectation. They showed that, due to the Chameleon effect,
the deviation of the non-linear power spectrum from GR is suppressed on small scales. It was also shown
that the fitting formula developed in GR such as Halofit \cite{Smith:2002dz} failed to describe this recovery
of GR and it overestimated the deviation from GR. Refs.~\cite{Oyaizu:2008sr, Oyaizu:2008tb, Schmidt:2008tn} used a multi-grid technique to solve the non-linear
equation for the scalar field. To make predictions for weak lensing, it is
necessary to model the power spectrum on smaller scales. Ref.~\cite{Koyama:2009me} derived a fitting
formula for the non-linear power spectrum in the so-called Post Parametrised Friedmann (PPF) framework
\cite{Hu07} based on perturbation theory and N-body simulations. Ref.~\cite{Beynon:2009yd}
calculated weak lensing signals for the future experiments using this fitting formula.
It was shown that the constraints on the model crucially depend on the modeling of
the non-linear power spectrum.

The resolution of N-body simulations performed in Refs.~\cite{Oyaizu:2008sr, Oyaizu:2008tb, Schmidt:2008tn}
is limited to $k <$ a few h Mpc$^{-1}$ for the power spectrum mainly due to that fact that
they used a fixed grid size. In this paper, we exploit a technique to recursively
refine a grid to solve the scalar field equation and the Poisson equation aiming to probe the power
spectrum down to $k \sim 20$ h Mpc$^{-1}$. We modify the publicly available N-body simulation code
{\tt MLAPM} and add a scalar field solver. We also study properties of halos in our simulations.
The Chameleon mechanism depends on local densities thus its effect depends on
the mass of halos. It is important to quantify the effectiveness
of the Chameleon mechanism to maximise our ability to distinguish between models.

This paper is organised as follows. In section II, we describe our $f(R)$ gravity models and present basic
equations to be solved.
In section III, we explain the details of our N-body simulations and how we analyse the simulation data
to obtain the power spectrum and halo properties. In section IV, we present the matter power spectrum
and properties of halos. We first use a conservative cut-off for the matter power spectrum to check the
consistency of our results with linear predictions on large scales and previous results in
Refs.~\cite{Oyaizu:2008sr, Oyaizu:2008tb, Schmidt:2008tn}. Then using the high resolution
simulations, we present the power spectrum on small scales and develop the fitting formula that captures
the effect of the Chameleon effects based on the PPF formalism. We then study the properties of halos by
measuring the mass functions and halo density profiles. We also measure the profiles of the scalar field
inside halos to quantify the effects of the Chameleon mechanism.
Section V is devoted to conclusions.

\section{$f(R)$ Gravity and Chameleon}

\label{sect:model}

In this section we briefly summarise the main ingredients of the $f(R)$ gravity theory and its properties, which are essential for the simulations and discussions below.

\subsection{The $f(R)$ Model}

A possible generalisation of GR is to add a term, which is a function of the scalar curvature $R$, to the Einstein-Hilbert action $S$ (see \cite{Sotiriou:2008rp, DeFelice:2010aj} for a review and references therein)
\be\label{eq:action}
S=\int {\rd}^4x\sqrt{-g}\left[\frac{R+f(R)}{16\pi{G}}+\mathcal{L}_{\m}\right] ,
\ee
in which $G$ is the Newtonian constant and $\mathcal{L}_{\rm M}$ is the Lagrangian density for matter fields
(radiation, baryons and dark matter, which in this paper is assumed to be cold).
Taking a variation of the above action with respect to metric yields the modified Einstein equations
\be\label{eq:Eins}
G_{\ab}+X_{\ab}=8\pi{G}T_{\ab}.
\ee
Here the extra term $X_{\ab}$ denotes the modification of GR,
\be\label{eq:Xab}
X_{\ab}={\fR}R_{\ab}-\left(\frac{f}{2}-{\Box}{\fR}\right)g_{\ab}-D_{\alpha}D_{\beta}{\fR} ,
\ee
where $ \fR\equiv\frac{\rd f(R)}{\rd R} $ denotes an extra scalar degree of freedom, dubbed \emph{scalaron},
$D_{\alpha}$ denotes the covariant derivative with respect to metric and $\Box$ is the Laplacian operator.
Taking the trace of the modified Einstein equation Eq. (\ref{eq:Eins}) yields
\be
\label{eq:trace}
\Box\fR=\frac{\partial V_{\rm eff}}{\partial f_R}\equiv\frac{1}{3}(R-f_R R+2f-8\pi G\rho)=0,
\ee
which governs the time evolution and spatial configuration of $f_R$. Given the function form of $f(R)$, a complete set of equations
governing the dynamics of the model is obtained.

We shall work here in an almost Friedman-Robertson-Walker (FRW) universe with the line element given as,
\be
\label{eq:ds2}
{\rm d}s^2=a^2(\eta)\left\{[1+2\Phi(\vec{x},\eta)]{\rm d}\eta^2-[1-2\Psi(\vec{x},\eta){\rm d}\vec{x}^2]\right\},
\ee
in which $\eta$ is the conformal time, $a(\eta)$ is the scale factor, $\Phi$ is the gravitational potential and $\Psi$ is
the spatial curvature perturbation.
Subtracting off the background part of this equation, we obtain a dynamical equation for the perturbation of the scalaron,
\be\label{eq:scalaron}
\nabla^2\dfR =-\frac{a^2}{3}[\dr(\fR)+8\pi{G}\delta\rho_{\m}],
\ee
where $\dfR=\fR(R)-\fR(\bar{R}),\dr=R-\bar{R},\delta\rho_{\m}=\rho_{\m}-\bar{\rho_{\m}}$, $\nabla$ is the
three dimensional gradient operator and we work in the quasistatic limit,
meaning that we assume the spatial variation of the scalaron is much larger than the time variation, so that we could neglect the time directives of $\fR$
and approximate $\Box\dfR$ as $\nabla^2\dfR$ \footnote{The validity of this assumption has been verified by Ref. \cite{Oyaizu:2008sr}, namely, they found that the amplitude of the 2-norm of the spatial derivative
is larger than that of the time derivative by a factor of $10^5$ at all redshifts. We also checked the validity of the
quasistatic assumption with our simulations, and found a
good consistency with Ref. \cite{Oyaizu:2008sr}.}.
Here the quantities with a bar denotes those evaluated in the cosmological background.
Notice that $\delta f_R, \delta R$ and $\delta\rho_{\rm M}$ are not necessarily small, and we call them perturbations just for convenience.

We can also obtain the counterpart of Poisson equation for the gravitational potential $\Phi$ by adding up the $00$ and $ii$ components of the modified Einstein equation Eq.~(\ref{eq:Eins}) (see Appendix A of \cite{Li:2010st} for some useful expressions using the above metric convention)
\be
\label{eq:Poisson}
 \nabla^2 \Phi=\frac{16\pi{G}}{3}a^2\delta\rho_{\m}+\frac{a^2}{6}\dr(\fR),
\ee
in which we have neglected terms such as $\ddot{\Phi}$ and $\frac{\dot{a}}{a}\dot{\Phi}$ as we are working in the quasistatic limit, and used Eq.~(\ref{eq:scalaron}) to eliminate $\nabla^2 \dfR$.

Given the density field, a functional form of $f(R)$ and the knowledge of the background evolution, Eqs. (\ref{eq:scalaron}) and (\ref{eq:Poisson}) are closed
for the scalaron and the gravitational potential. These are thus the starting point for our $N$-body simulations.

Note that the remaining modified Einstein equations give a relation between two metric perturbations $\Phi$ and $\Psi$
\be
\label{eq:fRaniso}\nabla^2(\Psi-\Phi)=\nabla^2\dfR,
\ee
where we assumed $\bar{f}_R \ll 1$ in the background.
Then combining Eqs. (\ref{eq:scalaron}), (\ref{eq:Poisson}) and (\ref{eq:fRaniso}), we find that
the relationship between the lensing potential $\Psi+\Phi$ and the matter overdensity
is the same in both GR and $f(R)$ scenarios, which is,
\be \label{eq:lenspot}
\nabla^2 (\Psi + \Phi) = 8 \pi G a^2 \delta \rho_{\rm M}.
\ee

\subsection{The Chameleon mechanism}

It is well known that $f(R)$ gravity models can incorporate the so-called Chameleon mechanism \cite{Khoury:2003rn, Mota:2006}, which are vital for their cosmological and physical viability \cite{Li:2007, Hu:2007nk}. The Chameleon mechanism was first proposed in the context of coupled scalar field theories, and used to give the scalar field an environment-dependent effective mass $m_\varphi$ so that $m_\varphi$ is very heavy in dense regions and thus the scalar-mediated fifth force is suppressed locally so as to avoid conflicts with experiments and observations. Because the $f(R)$ gravity is equivalent to a coupled scalar field theory in the Einstein frame with the functional form of $f(R)$ related to the potential of the scalar field, a similar mechanism is needed to ensure that the fifth force in $f(R)$ gravity is well within the limits set by experiments.

To see how the Chameleon mechanism works in $f(R)$ gravity models, let us consider their difference from GR. For Eq.~(\ref{eq:scalaron}), $\dfR$ vanishes identically in GR, and we have $\delta R = -8\pi G\delta\rho_{\rm M}$ which, substituted into Eq.~(\ref{eq:Poisson}), recovers the usual Poisson equation for GR,\be\nabla^2\Phi=4\pi Ga^2\delta\rho_{\rm M}.\ee Therefore, to make Chameleon mechanism work, we have to ensure that in Eq.~(\ref{eq:scalaron}) $\nabla^2\dfR=\nabla^2f_R$ is close to zero\footnote{This means that $\dfR$ is only sensitive to the local matter density distribution, which is merely a reflection of the fact that the extra scalar degree of freedom $\dfR$ has a heavy mass (in dense regions) and cannot propagate away from the source (and as such the fifth force is severely suppressed).}, meaning that $\fR$ must depend on $R$ very weakly (\ie,  be nearly constant) in high curvature region. Furthermore, if $\fR$ deviates significantly from zero, then Eq.~(\ref{eq:Xab}) means that the theory will become one with a modified Newtonian constant, which is not of our interest. A necessary condition for the Chameleon mechanism to work here is thus $|\fR|\ll1$ at least in the high curvature region\footnote{There is also requirement on the sign of $\fR$ to avoid instability in the perturbation growth \cite{Song:2007, Li:2007}.}.

The functional form of $f(R)$ must be carefully designed so that it can give rise to the late time cosmic acceleration of the universe without any conflict with the solar system tests by virtue of the Chameleon mechanism. One such model was studied in \cite{Li:2007}, where $f(R)\propto(-R)^n$ with $n\ll1$ ($n=0$ corresponds to a cosmological constant). A more interesting model was proposed in \cite{Hu:2007nk}, namely,
\be f(R)=
-m^2\frac{c_1(-R/m^2)^n}{c_2(-R/m^2)^n+1},
\ee
where $m^2\equiv8\pi
G\bar{\rho}_{\rm M,0}/3=H_0^2\Omega_{\rm M}$; note the minus sign in front of $R$ is due to our different sign convention from \cite{Hu:2007nk}. To evade the solar system test, the absolute value of the scalaron today, $|f_{R0}|$, should be sufficiently small. However the constraint is fairly weak ($|f_{R0}|<0.1$) \cite{Hu:2007nk} and it is satisfied in our models. In this scenario, the scalaron always sits at the minimum of the effective potential defined in Eq (\ref{eq:trace}), thus,
\be-\label{R}
\bar{R}\approx8\pi G{\bar{\rho}_{\rm
M}}-2\bar{f}=
3m^2\Big(a^{-3}+\frac{2}{3}\frac{c_1}{c_2}\Big).
\ee
To match the
$\Lambda$CDM background evolution, we need to have
\be
\frac{c_1}{c_2}=6\frac{\Omega_{\Lambda}}{\Omega_{\rm M}}.
\ee
Plugging
in the numbers $\Omega_{\Lambda}\approx0.76,\Omega_{\rm
M}\approx0.24$ into Eq~(\ref{R}), we have
$-\bar{R}\approx41 m^2\gg m^2$, which can be used to simplify the expression for the
scalaron,
\be\label{fR}
f_R=-\frac{c_1}{c_2^2}\frac{n(-R/m^2)^{n-1}}{[(-R/m^2)^{n}+1]^2}\approx-\frac{nc_1}{c_2^2}\Big(\frac{m^2}{-R}\Big)^{n+1}.
\ee
Therefore we can see that the two free model parameters are $n$ and $c_1/c_2^2$, and the latter is related to the
value of the scalaron today via
\be\label{eq:fR0}
\frac{c_1}{c_2^2}=-\frac{1}{n}\Big[3\Big(1+4\frac{\Omega_{\Lambda}}{\Omega_{\rm
M}}\Big)\Big]^{n+1}f_{R0},
\ee
We shall concentrate on the models with $n=1$ and $|f_{R0}|=10^{-4},10^{-5},10^{-6}$ in this paper, as in \cite{Oyaizu:2008sr}.

In the linear regime where $\delta \rho_{\rm M}/\bar{\rho}_{\rm M} \ll 1$,
we can linearised the scalaron equation by linearsing Eq.~(\ref{R}) with
respect to a cosmological background
\be
\label{eq:deltafR}
\delta f_R= - (n+1) \bar{f}_{R0} \left( \frac{\bar{R}_0}{\bar{R}} \right)^{n+1} \frac{\delta R}{\bar{R}},
\ee
where $\bar{R}$ is the background curvature and the subscript 0 implies that the quantity is evaluated today.
Then the scalaron equation (\ref{eq:scalaron}) becomes
\be
\label{eq:linearscalaron}
\nabla^2 \delta f_R= a^2 \bar{\mu}^2 \delta f_R - \frac{8 \pi G}{3} a^2 \delta \rho_{\rm M},
\ee
where
\be
\bar{\mu} = \lambda_c^{-1}= \left( \frac{1}{3 (n+1)} \frac{\bar{R}}{|\bar{f}_{R0}|} \left(
\frac{\bar{R}}{\bar{R}_0}\right)^{n+1} \right)^{1/2}.
\ee
In Fourier space, the solution can be easily obtained as
\be
\label{eq:dfRfourie}
\delta f_R = \frac{1}{3} \frac{8 \pi G a^2 \delta \rho_{\rm M}}{k^2 + a^2 \bar{\mu}^2}.
\ee
Then it is easy to show that we recover GR above the Compton wavelength $a/k > \lambda_c$ as
$\delta f_R$ is suppressed by the mass term compared with the gravitational potential, $|\delta f_R| \ll \Phi$.
This is because the scalaron cannot propagate beyond the Compton wavelength.
Below the compton wavelength, $\delta f_R  = - 8 \pi G \delta \rho_{\rm M}/3$ and we get
\be
\nabla^2 \Phi = \frac{16 \pi G}{3}  a^2 \delta \rho_{\rm M},
\ee
which implies that the gravitational constant is enhanced by $4/3$. This leads to the scale-dependent enhancement
of the linear growth rate. This linearisation around the cosmological background fails when $\delta f_R$ becomes
larger compared with the background field $\bar{f_R}$, $\delta f_R \gg |\bar{f}_{R}|$.
This happens in the dense region because $\delta f_R$ is driven by the matter perturbations $\delta \rho_{\rm M}$.
In these dense regions, the curvature is large and Eq.~(\ref{eq:deltafR}) ensures that $f_R$ is suppressed and GR is recovered,
realising the Chameleon mechanism. In Fig.~1, we plot the time evolution of the Compton wavelength and the background
$f_R$ field with $n=1$ and $|f_{R0}|=10^{-4},10^{-5},10^{-6}$, which is useful to understand the recovery of GR
in our simulations.

\begin{figure}[t]
\includegraphics[scale=0.2]{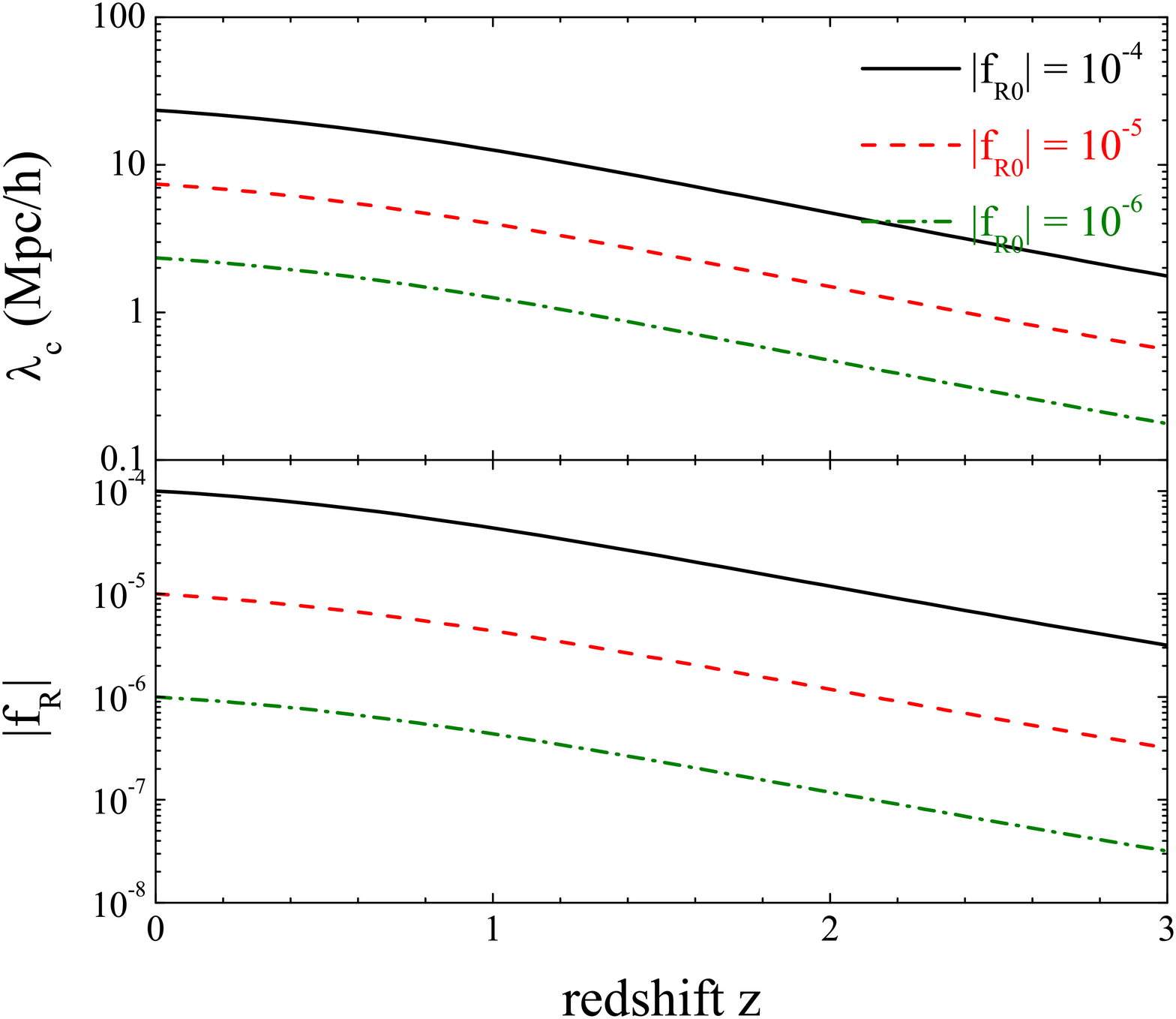}
\caption{The evolution of the Compton wavelength $\lambda_{c}$ (upper panel) and the absolute value of the background scalar field $|f_R|$ as a function of redshift $z$. The $f(R)$ models of $|f_{R0}|=10^{-4},10^{-5},10^{-6}$ are illustrated using black solid, red dashed and green dash-dot curves respectively.}\label{fig:comp-sf}
\end{figure}

\section{The $N$-body Simulations}

\label{sect:Nbody}

In the model of \cite{Hu:2007nk}, the fifth force due to the propagation of $\dfR$ is suppressed by many orders of magnitudes than gravity only in very dense regions, while in less dense regions it has similar strength as gravity. Such a dramatic change of the amplitude of the fifth force across space indicates that Eq.~(\ref{eq:scalaron}) should be highly nonlinear, which the readers can convince themselves by appreciating the nonlinearity of $\fR$ [cf.~Eq.~(\ref{fR})]. Consequently treatments involving linearisation fails to make correct predictions, and $N$-body simulations which explicitly solve the nonlinear equation for $\dfR$ are needed.

In a series of papers \cite{Oyaizu:2008sr, Oyaizu:2008tb, Schmidt:2008tn}, Oyaizu \etal~pioneered in this direction, by solving Eq.~(\ref{eq:scalaron}) on a regular mesh to compute the total force on particles. Their results show some interesting and cosmologically testable predictions of the $f(R)$ gravity. However, the resolution of their simulations was limited by their regular mesh. Here, we perform similar simulations, but with an adaptive grid which self-refines in the high density regions. This technique has been applied previously in \cite{Li:2009sy, Li:2010mq, Li:2010nc} and works well.

As we are interested in how well the Chameleon mechanism actually works, besides the full $f(R)$ simulations, we also perform some non-Chameleon runs with the Chameleon effect artificially suppressed (see below). For clarity we shall refer to these two classes of simulations respectively as full $f(R)$ and non-Chameleon simulations.

\subsection{Outline of the Simulation Algorithm}

We have modified the publicly available $N$-body simulation code {\tt MLAPM} \cite{Knebe:2001}, which is a self-adaptive particle mesh code, for our $N$-body simulations. It has two sets of meshes: the first mesh includes a series of increasingly refined regular meshes covering the whole cubic simulation box, with respectively $4, 8, 16, \cdots, N_d$ cells on each side, where $N_d$ is the size of the domain grid, which is the most refined of these regular meshes. This set of meshes are needed to solve the Poisson equation using multigrid method or fast Fourier transform (for the latter only the domain grid is necessary). When the particle density in a cell exceeds a predefined threshold, the cell is further refined into eight equally sized cubic cells; the refinement is done on a cell-by-cell basis and the resulted refinement could have arbitrary shape which matches the true equal-density contours of the matter distribution. This second set of meshes are used to solve the Poisson equation using a linear Gauss-Seidel relaxation.

Our core change to the {\tt MLAPM} code is the addition of the routines which solve Eq.~(\ref{eq:scalaron}), which is similar to that in \cite{Li:2010mq, Li:2010nc}. These include:
\begin{enumerate}
\item We have added a solver for $\dfR$, based on Eq.~(\ref{eq:scalaron}), which uses a nonlinear Gauss-Seidel relaxation iteration and the same convergence criterion as the default Poisson solver of {\tt MLAPM}. It adopts a V-cycle instead of the default self-adaptive scheme in arranging the Gauss-Seidel iterations, because the latter is found to be plagued by the problem of oscillations between adjacent multigrid levels, and the convergence is too slow.
\item The value of $\dfR$ solved from the above step is used to complete the computation of the source term for the Poisson equation Eq.~(\ref{eq:Poisson}), which is solved using a fast Fourier transform on the regular domain grids and Gauss-Seidel relaxation on the irregular refinements.
\item The value of $\Phi$ solved from the above step is used to compute the total force on the particles by a finite difference, and then the particles are displaced and accelerated as in usual $N$-body simulations, according to
\begin{eqnarray}
\frac{d{\bf x}_c}{dt_c} &=& \frac{{\bf p}_c}{a^2},\\
\frac{d{\bf p}_c}{dt_c} &=& -\frac{1}{a}\nabla_c\Phi_c,
\end{eqnarray}
\end{enumerate}
where a subscript $_c$ denotes code unit.
More details about the code could be found in \cite{Knebe:2001, Li:2010nc}.

\subsection{$N$-body Equations in Code Units}

To implement Eqs.~(\ref{eq:scalaron}, \ref{eq:Poisson}) into our numerical code, we have to rewrite them using code units, which are given by
\ba \mathbf{x}_c\ =\ \frac{\mathbf{x}}{B},\ \ \ \mathbf{p}_c\ =\ \frac{\mathbf{p}}{H_0{B}},\ \ \ t_c\ =\ tH_0,\nonumber\\
\Phi_c\ =\ \frac{\Phi}{(H_0{B})^{2}},\ \ \ \rho_c\ =\ \frac{\rho_{\rm M}}{\bar{\rho}_{\rm M}},\ \ \ \nabla_c\ =\ B\nabla,
\ea
where $B$ is the size of the simulation box and $H_0=100h$~km/s/Mpc. We also need to choose a code unit for $\dfR$ (or $\fR$), but this is done in different ways for the Chameleon and non-Chameleon simulations for convenience, and here we shall discuss them separately.

\subsubsection{The Full $f(R)$ Simulations}

In our full $f(R)$ simulations, the quantity $\delta R$ in Eq.~(\ref{eq:scalaron}) is written in its exact form $\delta R  = R-\bar{R}$ with
\begin{eqnarray}\label{eq:R}
R &=& -m^2\left(-n\frac{c_1}{c_2^2}\frac{1}{\fR}\right)^{\frac{1}{n+1}},\\
\bar{R} &=& -3m^2\left(a^{-3}+4\frac{\Omega_\Lambda}{\Omega_{\rm M}}\right),
\end{eqnarray}
where $\Omega_{m}$ and $\Omega_{\Lambda}$ are respectively the present fractional energy densities for matter and dark energy, and $m^2=\Omega_{\rm M}H_0^2$. With the aid of these, we can rewrite Eq.~(\ref{eq:scalaron}) using code units as
\begin{eqnarray}
\label{eq:dfRcode}\frac{ac^2}{\left(BH_0\right)^2}\nabla^2\fR &=& -\Omega_{\rm M}(\rho_c-1)+\frac{1}{3}\Omega_{\rm M}a^3\left(-n\frac{c_1}{c_2^2}\frac{1}{\fR}\right)^{\frac{1}{n+1}}\nonumber\\
&&-\Omega_{\rm M}\left(1+4\frac{\Omega_\Lambda}{\Omega_{\rm M}}a^3\right).
\end{eqnarray}
Note that we will drop the subscript of $\nabla_c$ for simplicity from now on.

In the simulations, $|\fR|$ spans a range of many orders of magnitude, from $\sim10^{-13}$ in the very dense regions to $\sim10^{-5}$ in the low density regions. Obviously, the above equation is very sensitive to numerical errors in the trial solution to $\fR$, making it difficult to solve $\fR$ accurately. Furthermore, Eq.~(\ref{eq:R}) indicates that $\fR>0$ would lead to unphysical (imaginary) $R$, and if this happens by accident in the simulations due to the numerical
errors (which is very likely because $|\fR|\ll1$) then the computation cannot carry on further. To solve these problems, we introduce a new variable $u=\log(-\fR)$ (which is slightly different from the choice of \cite{Oyaizu:2008sr}). Then, typically $u\in(-30,-10)$, which does not vary much across the simulation box and $\fR=-e^u<0$ no matter which value $u$ takes.

Replacing $\fR$ with $u$, we can rewrite Eq. (\ref{eq:dfRcode}) as (after some rearrangement),
\begin{eqnarray}\label{eq:sf_codeunit_cham}
\frac{ac^2}{\left(BH_0\right)^2}\nabla^2e^u &=& \Omega_{\rm M}\rho_c + 4\Omega_\Lambda a^3\nonumber\\
&&-\frac{1}{3}\Omega_{\rm M}a^3\left(n\frac{c_1}{c^2_2}\right)^{\frac{1}{n+1}}e^{-\frac{u}{n+1}}.
\end{eqnarray}

Similarly, in terms of $u$ and using the code units, we can rewrite the Poisson equation as,
\begin{eqnarray}\label{eq:poi_codeunit_cham}
\nabla^2\Phi_c &=& 2\Omega_{\rm M}\rho_c - \frac{3}{2}\Omega_{\rm M} + 2\Omega_\Lambda a^3\nonumber\\
&&-\frac{1}{6}\Omega_{\rm M}\left(n\frac{c_1}{c_2^2}\right)^{\frac{1}{n+1}}e^{-\frac{u}{n+1}}a^3.
\end{eqnarray}

Eqs.~(\ref{eq:sf_codeunit_cham}, \ref{eq:poi_codeunit_cham}), after discretisation (Appendix~\ref{appen:discrete}), can be implemented in the modified {\tt MLAPM} code straightforwardly.

\subsubsection{The Non-Chameleon Simulations}

As mentioned above, because the Chameleon effect is one of the most important features of our $f(R)$ model, we also would like to see what happens if we artificially suppress it. Unlike \cite{Oyaizu:2008tb, Schmidt:2008tn}, in our simulations we achieve this by explicitly linearising the equation for $\fR$ without using the solution for $\delta f_R$ given in Eq. (\ref{eq:dfRfourie}). Since the Chameleon effect originates from the nonlinearity of $\fR$, the linearisation will largely weaken it, justifying the name non-Chameleon simulations.

The linearisation is easily done by simply setting $\delta R = \frac{\partial R}{\partial\fR}\dfR$, which changes Eq.~(\ref{eq:scalaron}) to
\be
c^2\nabla^2\fR = c^2\nabla^2\dfR = -\frac{8\pi G}{3}\delta\rho_{\rm M}a^2 - \frac{1}{3}\frac{\partial R}{\partial\fR}\dfR a^2.
\ee
In this case, we find it more convenient to choose $\dfR$ as our variable, then using the relation
\be
\frac{\partial R}{\partial\fR} = -m^2\frac{c_2^2}{c_1}\frac{1}{n(n+1)}\left[3\left(a^{-3}+4\frac{\Omega_\Lambda}{\Omega_{\rm M}}\right)\right]^{n+2},
\ee
we can rearrange the above equation to obtain
\begin{eqnarray}\label{eq:sf_codeunit_nc}
&&\frac{ac^2}{\left(BH_0\right)^2}\nabla^2\dfR\nonumber\\ &=& -\Omega_{\rm M}(\rho_c-1)\nonumber\\
&&+\frac{1}{3}\frac{\Omega_{\rm M}a^3}{n(n+1)}\frac{c_2^2}{c_1}\left[3\left(a^{-3}
+4\frac{\Omega_\Lambda}{\Omega_{\rm M}}\right)\right]^{n+2}\dfR.
\end{eqnarray}
Note that after the linearisation the mass of the scalaron depends only on the background matter density; but if $|f_{R0}|$ is chosen to be small enough (such that $c_2^2/c_1\gg1$) the mass can be very heavy, and thus the fifth force still gets suppressed, even though the suppression is not as strong as in the full $f(R)$ simulations. Furthermore, the suppression of the fifth force in this case is universal and independent of the local matter density, again justifying its name -- non-chameleon model.

Similarly, the modified Poisson equation appears to be
\begin{eqnarray}\label{eq:poi_codeunit_nc}
&&\nabla^2\Phi_c\nonumber\\ &=& 2\Omega_{\rm M}(\rho_c-1)\nonumber\\
&&-\frac{1}{6}\frac{\Omega_{\rm M}a^3}{n(n+1)}\frac{c_2^2}{c_1}
\left[3\left(a^{-3}+4\frac{\Omega_\Lambda}{\Omega_{\rm M}}\right)\right]^{n+2}\dfR.
\end{eqnarray}

Eqs.~(\ref{eq:sf_codeunit_nc}, \ref{eq:poi_codeunit_nc}), after discretisation (Appendix~\ref{appen:discrete}), can be implemented in the modified {\tt MLAPM} code directly.

\subsection{Code Tests}

As the major modification to the {\tt MLAPM} code is for the $\dfR$ solver, we have to check carefully whether it works correctly. For this purpose we have solved Eq.~(\ref{eq:sf_codeunit_cham}) around a point-like mass \cite{Oyaizu:2008sr} and compared the result with the analytic solution; we find that the numerical and analytic solutions agree as well as in \cite{Oyaizu:2008sr} (Fig.~2 there). To avoid repetition we shall not display the results here.

Other, indirect, tests of our code include comparing the predicted matter power spectrum to the linear perturbation results to see whether they agree on large scales. We shall discuss these in turn as the paper unfolds.

\subsection{Simulation Details}

\begin{table}
\begin{tabular}{c|cccc}
  \hline\hline
                &               &    Box size (Mpc/h)            &\\
                & $256$ &  $128$  & $64$  \\
                \hline
  $N_{\rm sim}$ & 10 & 10 & 10  \\
  $N_{\rm p}$ & $256^3$ & $256^3$ & $256^3$  \\
  $N_{\rm grid}$ & 128 & 128 & 128  \\
  $k_{N/2} $(h/Mpc) & $0.79$ & $1.57$ & $3.14$  \\
  $k_{\ast}$(h/Mpc) & 5.5 & 11.0 & 22.0  \\
  Refinement levels & 8 & 9 & 10  \\
  Force resolution (Kpc/h) & 12 & 23 & 94  \\
  Mass resolution ($10^{11}M_{\odot}$/h)& 13.3 & 1.75 & 0.21  \\

  \hline\hline
\end{tabular}
\caption{Technical indices for the simulations presented in this work. The quantity $N_{\rm sim}$ denotes the number of realisations we run for each model, $N_{\rm p}$ is the total
number of particles, $N_{\rm grid}$ is the number of domain grids one each side, and $k_{N/2}$ and $k_{\ast}$, which are defined in Eq. (\ref{eq:kcutoff}), show the half Nyquist scale and the $k$-cutoff we actually used in this work respectively.}\label{tab:nbody}
\end{table}

We choose to run our simulations in a cosmology that is consistent with the WMAP observations. Specifically, we parametrise the universe in the form of
\be
\textbf{p}
\equiv\{\mathcal{C},\mathcal{G}\}.
\ee
The symbols $\mathcal{C}$ and $\mathcal{G}$ denote the subsets of cosmological and $f(R)$ parameters, respectively.
\be
\label{eq:cosmoparam}
\mathcal{C}\equiv\{\Omega_{\rm b}h^2,\Omega_{\rm c}h^2,\Omega_K,H_0,n_s,\sigma_8\},
\ee
where $\Omega_{\rm b}h^2$ and $\Omega_{\rm c}h^2$ denote the physical energy density for baryons and cold dark matter respectively, $\Omega_K$ is the curvature, $H_0$ is the Hubble constant today, $n_s$ stands for the index of the primordial power spectrum at a pivot scale of $k=0.05$ Mpc$^{-1}$, and $\sigma_8$ measures the amplitude of the (linear) power spectrum on the scale of $8$ Mpc/h.
\be
\label{eq:MG}
\mathcal{G}\equiv\{n,f_{R0}\}.
\ee
Here $n$ and $f_{R0}$ specify a $f(R)$ model as described in Sec. \ref{sect:model}.
We set the cosmological parameters as $\mathcal{C}=\{0.04181,0.1056,0,73,0.958,0.8\}$, while for the $f(R)$ parameters, we fix $n=1$ and simulate for three models with $f_{R0}=-10^{-4},-10^{-5},-10^{-6}$. We also run the standard GR model ($f_{R}=0$) for the purpose of comparison. For each $f(R)$ model, we also run the simulation without the Chameleon mechanism using the same initial condition and background evolution. Since we will refer to these simulations frequently later in the text, we assign them abbreviations to lighten the notations, as listed below,

\begin{itemize}
  \item F4 = Full $f(R)$ simulation with $|f_{R0}|=10^{-4}$;
  \item F5 = Full $f(R)$ simulation with $|f_{R0}|=10^{-5}$;
  \item F6 = Full $f(R)$ simulation with $|f_{R0}|=10^{-6}$;
  \item N4 = non-Chameleon simulation with $|f_{R0}|=10^{-4}$;
  \item N5 = non-Chameleon simulation with $|f_{R0}|=10^{-5}$;
  \item N6 = non-Chameleon simulation with $|f_{R0}|=10^{-6}$;
  \item GR = Simulation for a $\Lambda$CDM model in GR.
\end{itemize}

For each model listed above, we run 10 simulations with 10 different initial conditions (ICs), which we call 10 `realisations', to reduce the sample variance. To extend the range of scales, we run these simulations in 3 difference boxes with size $B=256,128$ and $64$ Mpc/h. The technical details are summarised in Table \ref{tab:nbody}. We generate the ICs using {\tt Grafic}, an IC generator included in the {\tt COSMICS} package \cite{Bertschinger:1995er}, at redshift $z=49$ in GR, and then evolve the system using our modified version of \texttt{MLAPM} for $f(R)$, as well as using the default \texttt{MLAPM} for GR simulations.

\subsection{Data Analysis}

Data analysis is of great significance for the simulation. In this section, we will detail our pipeline to obtain the snapshots, matter power spectra, mass function and the density and scalar field profiles out from the raw simulation output.

\subsubsection{Snapshots}

Visualisations of the simulation result is helpful to understand the physics in an intuitive way. For this purpose, we will show the 2-D snapshots for the distribution of overdensity $\delta\equiv\delta\rho/\rho$, the perturbation of the scalar field $\dfR$ and gravitational potential $\Phi$. For the visualisation, we output the data for ln$(1+\delta)$, $-2\dfR$ and $\Phi$ on a $400\times400\times400$ grid from our simulation, and project the 3-D volume onto a 2-D plane to make image snapshots. The resolution we use here is much lower than that we use for calculation in the code, but it is sufficient for the purpose of visualisation.

\subsubsection{Binning, Average and Variance}

We will present our simulation results (power spectra, mass function and profiles) in terms of data bins along with error bars. Suppose our observable is called $O$ which is a function of $x$, and we have $N_{\rm raw}$ raw samples measured from simulation for $O$ in some range of $x$ that we are interested in, then we take an average of $O$ over 10 realisations first, make logarithmic bins in $x$, and then calculate the mean value $\bar{O_i}$ and error bar $\sigma(O_i)$ for the $i$th bin via,
\be
\label{eq:errbar}
\bar{O_i}=\sum_{j=1}^{N_i} O_{ij}/N_i,~~\sigma^2(O_i)=\sum_{j=1}^{N_i} (O_{ij}-\bar{O_i})^2/N_i,
\ee
where $N_i$ is the total number of raw samples falling into the $i$th bin.

\subsubsection{Matter Power Spectrum}

The matter power spectrum $P(k)$ is an important statistical measure of the matter clustering on different scales in Fourier space.  We measure the matter power spectrum using a tool called \texttt{POWMES} \cite{Colombi:2008dw}. \texttt{POWMES} estimates the Fourier modes of a particle distribution based on a Taylor expansion of the trigonometric functions, and it is able to
accurately determine and correct for the biases induced by the discreteness and by the truncation of the Taylor expansion. Also, the aliasing effect is safely negligible if the order of Taylor expansion $N=3$, which we chose to use. One could even accurately measure $P(k)$ up to the scale of $k\sim1024\pi/$(Box size) with the `folding' operation, which we adopted in our analysis.

We firstly measure the power spectrum for GR simulations using \texttt{POWMES}, and use $P_{\rm GR}$ as the observable $O$. Then we utilise Eq. (\ref{eq:errbar}) to estimate the mean value and error bar for $P_{\rm GR}(k)$ for each $k$ bin and compare it to the Halofit (Smith \etal, Ref. \cite{Smith:2002dz}) prediction. Then we measure the power spectra for full $f(R)$ and non-Chameleon simulations, use the relatively difference
\be
\label{eq:dPoP}
\Delta{P}/P_{\rm GR}{\equiv}(P_{f(R)}-P_{\rm GR})/P_{\rm GR},
\ee
as the observable $O$ and similarly use Eq. (\ref{eq:errbar}) to get estimates of $\Delta{P}/P_{\rm GR}$.

\subsubsection{Halo Mass Function}

We use the \texttt{MHF} tool \cite{Gill:2004km} (the Halo-finder for \texttt{MLAPM}) to resolve halos in our simulation. The halo-finding algorithm we use is similar to that in Ref. \cite{Schmidt:2008tn} -- we assign the particles to the grids using a
Triangular Shaped Cloud (TSC) interpolation, and count the particles within a sphere around the highest overdensity grid point. We increase the radius of the sphere until the overdensity $\delta$ reaches a threshold $\Delta$. This process can be repeated until all the halos are identified, and we only keep those that are made of at least $800$ particles for our analysis.
In this paper we choose $\Delta$ to be the virial overdensity in $\Lambda$CDM model, $\Delta=\Delta_{\rm vir}$.
Thus the mass is defined by $M= 4 \pi \Delta_{\rm vir} r_{\rm vir}^3/3$ where $r_{\rm vir}$ is the radius when the
overdensity reaches the threshold $\delta = \Delta_{\rm vir}$. $\Delta_{\rm vir}$ depends on redshifts; $\Delta_{\rm vir} = 373.76$ at $z=0$ and $\Delta_{\rm vir} = 242.71$ at $z=1$ in our GR models. We use the same definition of the mass in $f(R)$ simulations in order to make a fair comparison between $f(R)$ gravity models and GR models. However, we should bear in
mind that in $f(R)$ gravity the virial overdensity is different from that in GR \cite{Schmidt:2008tn} and a care must
be taken when we compare the mass function in our $f(R)$ simulations to observations.

In this work we use the definition of the halo mass function as the halo number density per logarithmic interval in the virial mass $M$ in GR, \ie,
\be
\label{eq:massfunc}
n_{{\rm ln}{M}}\equiv\frac{\rm d{n}}{{\rm d ln}{M}}.
\ee
This definition is different from Ref. \cite{Schmidt:2008tn} where they use $\Delta = 300$ to define the mass. Thus
a direct comparison is not possible without rescaling the mass.

Since the halos in $f(R)$ and GR simulations have different number and mass in general \footnote{The number and mass of the halos are in general different even for the same gravity model for different realisations.}, and we are interested in the relative difference of the mass function in $f(R)$ model with respect to that in GR, we need to make sure that we are comparing the same quantity in different gravity models. Therefore, for the simulations with the same box size, we find the overlapping halo mass range for full $f(R)$, non-Chameleon and GR simulations for all the 10 realisations, and make logarithmic bins in mass in this range. Then we count the number of halos falling into each mass bin for $f(R)$ and GR cases respectively, and calculate the relative difference $\Delta n_{{\rm ln}{M}}/n_{{\rm ln}{M}}^{\rm GR}\equiv(n_{{\rm ln}{M}}^{f(R)}-n_{{\rm ln}{M}}^{\rm GR})/n_{{\rm ln}{M}}^{\rm GR}$ for each bin for every realisation. Finally we average over 10 realisations to calculate the mean value and variance for each mass bin.

\subsubsection{Halo Profile}

We modified \texttt{MHF} to analyse the profile of the perturbation of the scalar field $\dfR$, and the gravitational potential $\Phi$ as well as the overdensity $\delta$ as a function of the rescaled virilised halo radius $r/r_{\rm vir}$. To see the relative difference of the density profile with respect to that in GR, we have to identify the same halos in the full, non-Chameleon and GR simulations for each realisation. Since more halos are produced generally in $f(R)$ models, we start from the GR halos in each realisation, and decide whether to include it for the analysis in the following steps,

\begin{enumerate}
  \item For the first realisation in GR, record the position and mass for the first halo, also the number of particles made up of this halo;
  \item For the same realisation for the 6 $f(R)$ simulations [F4,F5,F6 and N4,N5,N6], search for the same halo as it in GR. If all the following conditions are satisfied,
      \begin{enumerate}
        \item The number of particles made up of this halo should be greater than $800$;
        \item In all the 6 $f(R)$ simulations, a halo with the similar position (absolute distance to the corresponding halo in GR smaller than 0.1\% of the box size) and a similar mass (relative mass difference $|\Delta M/M-1|<1000\%$) is found;
       \end{enumerate}
      we assume it is the same halo as in GR simulations.
  \item Repeat the halo identification process so that all the halos with counterpart in $f(R)$ are selected;
  \item For every selected halo, rescale $r$ to $r/r_{\rm vir}$ and linearly interpolate the overdensity $\delta$ to obtain $\delta$ for the same $r/r_{\rm vir}$ for different gravity models;
  \item Calculate the relative difference $\Delta\delta/\delta_{\rm GR}\equiv(\delta_{f(R)}-\delta_{\rm GR})/\delta_{\rm GR}$ and feed $\Delta\delta/\delta_{\rm GR}$ into Eq. (\ref{eq:errbar}) to obtain the statistics for density profile.
\end{enumerate}

It is interesting to test the working efficiency of Chameleon mechanism inside the halos, and this can be realised by comparing the profiles of $\dfR$ to that of $\Phi$. If Chameleon does not work at all where the mass term of the scalar field can be ignored, \ie, $\delta R=0$, from Eqs. (\ref{eq:scalaron}) and (\ref{eq:Poisson}), we have
\be
-2\dfR=\Phi.
\ee
Therefore we define an efficiency parameter $\Gamma$ for the Chameleon mechanism as,
\be
\label{eq:gamma}
\Gamma\equiv|-2\dfR/\Phi.|
\ee
If $\Gamma\sim1$ then Chameleon hibernates, and if $\Gamma\ll1$ then it means Chameleon is very active. For the selected halos, we calculate $\Gamma$ for both full and non-Chameleon simulations and feed it into Eq. (\ref{eq:errbar}) to obtain the statistics for the profile of $\Gamma$ of the Chameleon mechanism.

\section{Results}

In this section, we will present our simulation results including the snapshots, matter power spectrum, the halo mass function and the profiles of the halo density and the efficiency parameter $\Gamma$.

\subsection{Snapshots}

\begin{figure*}[htp]
\includegraphics[scale=0.3]{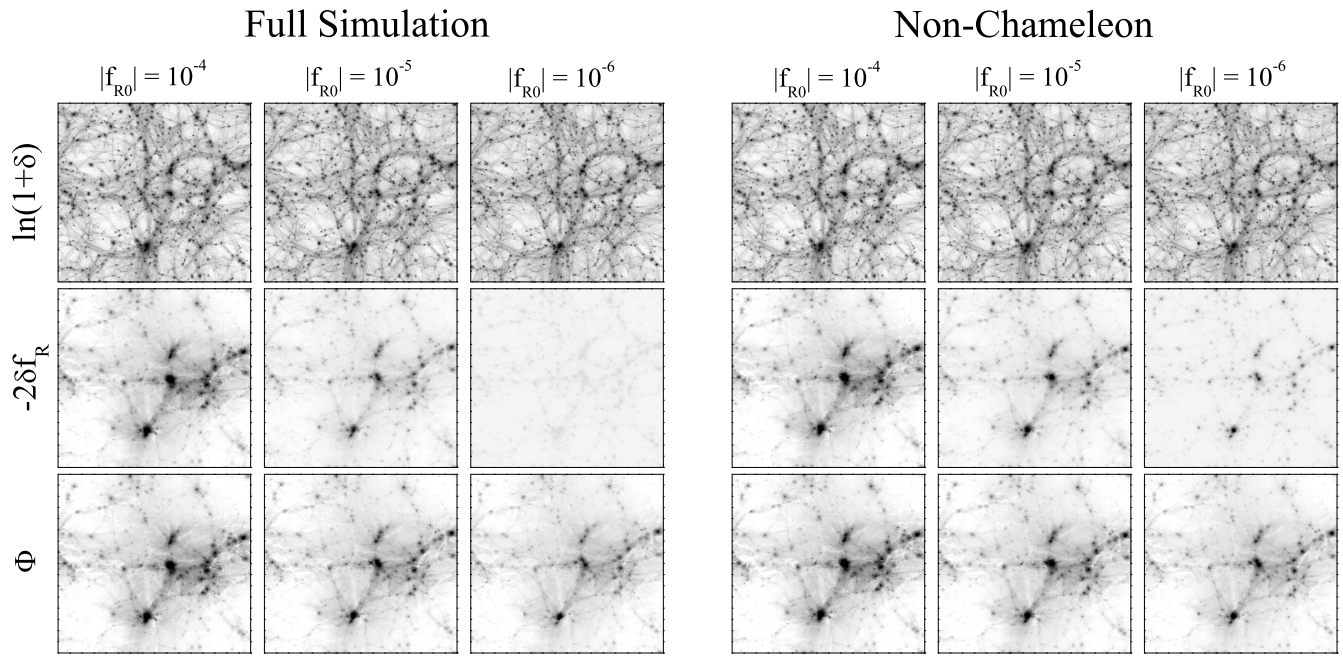}
\caption{The snapshots for density perturbation $\delta$, scalaron perturbation $\dfR$, and the gravitational potential $\Phi$ taken from the full $f(R)$ simulations (6 left panels) and the non-Chameleon simulations (6 right panels) for three values of $|f_{R0}|$, namely, $|f_{R0}|=10^{-4},10^{-5},10^{-6}$ from left to right. The box size of the simulation is $64$ Mpc/h, and all the snapshots are taken at $z=0$. Note that we use a same color scheme for all the plots of $-2\dfR$ and $\Phi$, namely, $-2\dfR,\Phi\in[-2.0,0.1]\times10^{-5}$, so that one could directly compare them to see the working efficiency of the Chameleon mechanism (see text for details). We use another color scheme for all the plots of the density distribution, say, ln$(1+\delta)\in[3.0,8.0]$.}\label{fig:snapshot}
\end{figure*}

\begin{figure*}[htp]
\includegraphics[scale=0.3]{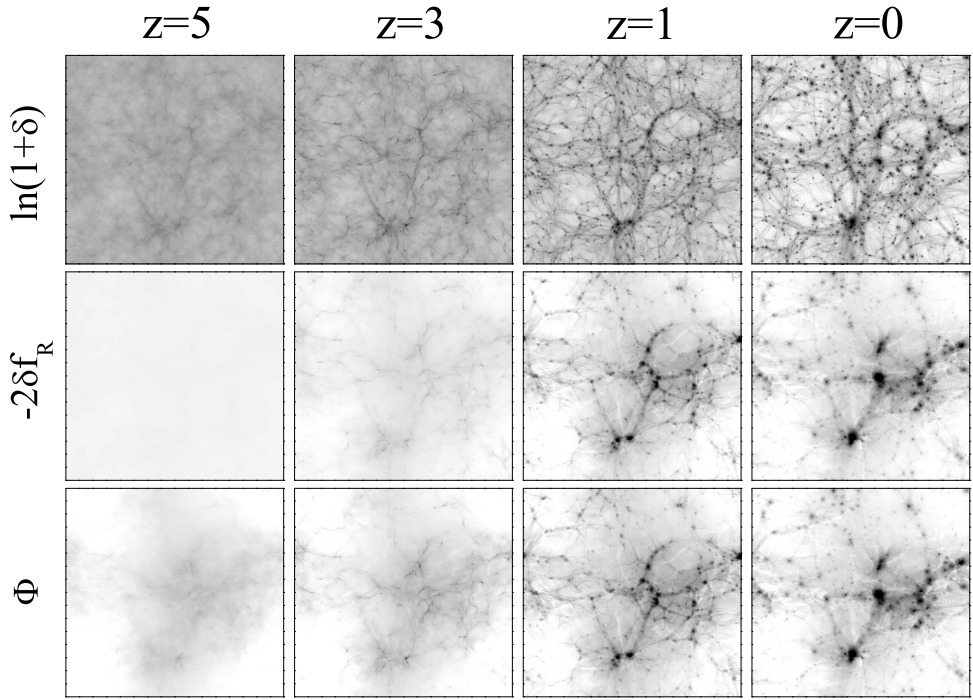}
\caption{The snapshots for density perturbation $\delta$, scalaron perturbation $\dfR$, and the gravitational potential $\Phi$ taken from the full $f(R)$ simulations with $|f_{R0}|=10^{-4}$ at four redshifts, namely, $z=5,3,1,0$ as illustrated in the figure. The simulation box size and color scheme are exactly the same as that in Fig. \ref{fig:snapshot}.}\label{fig:snapshot2}
\end{figure*}

First, let us view the snapshots first to get some basic idea on the physics of our simulations. In Fig. \ref{fig:snapshot}, we show the snapshots for the density perturbation $\delta$, the scalaron perturbation $\dfR$, and the gravitational potential $\Phi$ taken from the full $f(R)$ and non-Chameleon simulations for three values of $|f_{R0}|$. These snapshots are taken from one realisation with the box size $64$ Mpc/h at $z=0$. The density plots show the familiar web-like structures, while the scalar field and potential have similar pattern, but are smoother. From the snapshots we can see that $-2\dfR\sim\Phi$ for the F4, F5, N4 and N5 simulations, and the Chameleon works very well for F6 cases ($|\delta f_R|\ll|\Phi|$)
These are natural -- the Chameleon mechanism works when $\dfR > |\bar{f}_R|$ and $\dfR \leq \Phi \sim 10^{-5}$. Thus the F5 simulations are the critical case where the Chameleon mechanism is about to fail today.
Note that in the N6 simulation we also find that $|2\delta f_R|$ is smaller than $|\Phi|$, because here the scalaron also has
a heavy mass and a short Compton wavelength (though it is independent of local densities).

In Fig. \ref{fig:snapshot2}, we show the same quantities for the F4 case at higher redshifts. Interestingly, we can see that the Chameleon does work very well at $z \geq 3$, and it hibernates at $z < 3$. This is because the background field $|\bar{f}_R|$ is small at high redshifts (see Fig.~1) thus the condition for the Chameleon mechanism, $\dfR > |\bar{f}_R|$, is easily satisfied.

\subsection{Matter Power Spectrum}

In this section, we present the result for matter power spectrum in $f(R)$ theory, in comparison with that in GR.

\subsubsection{Tests for GR simulations}
\begin{figure}
\includegraphics[scale=0.2]{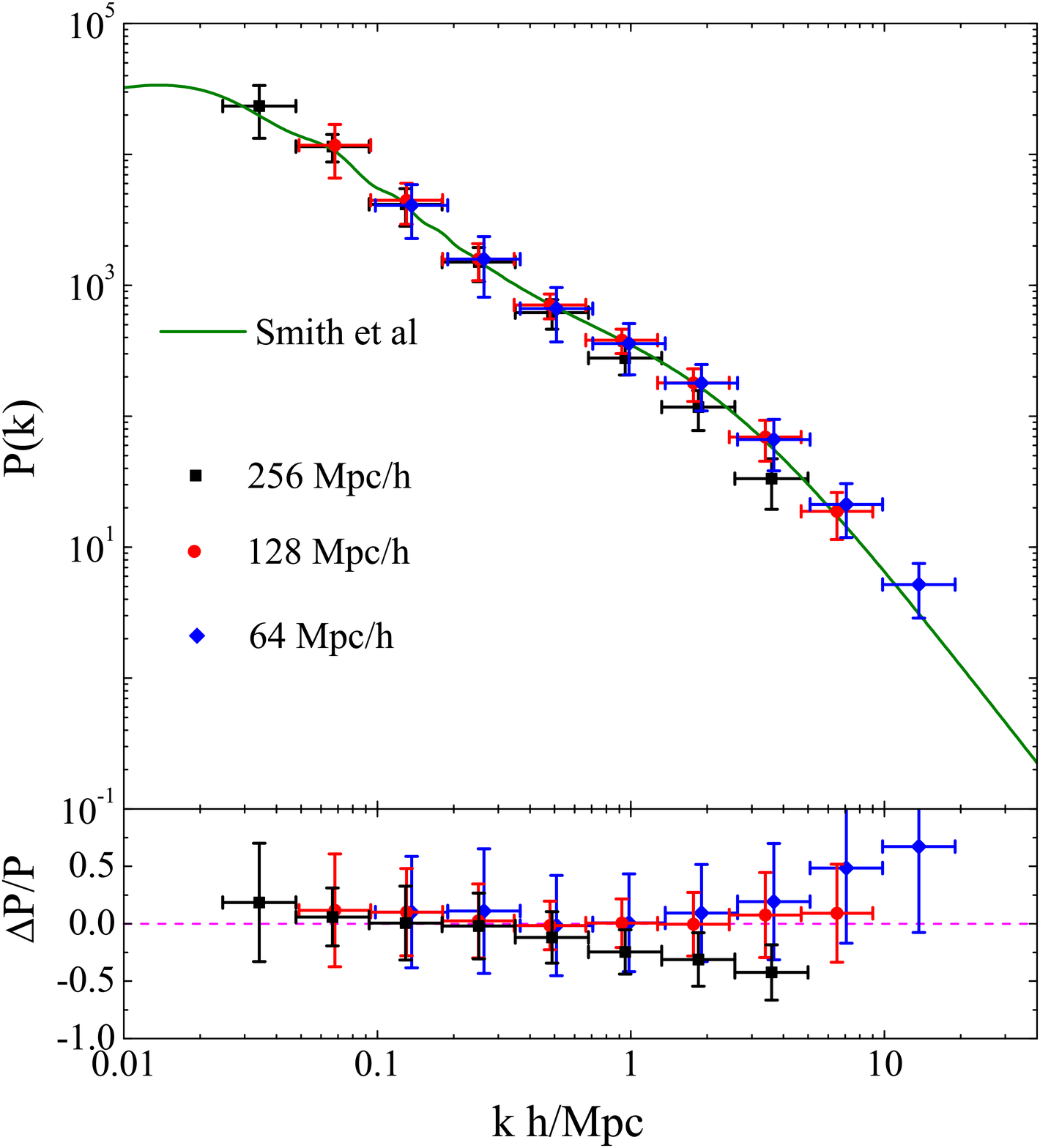}
\caption{Matter power spectra for standard GR at redshift $z=0$ (upper panel). The simulation results from different boxes are shown with error bars, while the Halofit result is over-plotted (green solid curve) for comparison. In the lower panel, the relative difference with respect to the Halofit result is shown, and the dashed magenta line illustrates $\Delta{P}=0$ to guide eyes.}\label{fig:pkGR}
\end{figure}

To start with, we test the accuracy of our GR simulations by comparing our results to the Halofit prediction \cite{Smith:2002dz}.
The result is shown in Fig. \ref{fig:pkGR}. As one can see, our $B=128$ Mpc/h simulation is very consistent with the Halofit prediction up to $k\sim10$ h/Mpc, namely, the relative difference is smaller than 10\% on all scales. The $B=256$ Mpc/h simulation tends to underestimate the power at $k\gtrsim0.4$ h/Mpc, which is due to the fact that we do not have enough particles to resolve small scales in a large box. The $B=64$ Mpc/h simulation shows much more power than that predicted by Halofit on scales $k\gtrsim5$ h/Mpc. This does not necessarily mean that our simulation is not accurate on those scales, instead, the Halofit prediction has never been tested on such small scales, thus it might be inaccurate.

\subsubsection{Low-resolution tests for $f(R)$}

\begin{figure}
\includegraphics[scale=0.2]{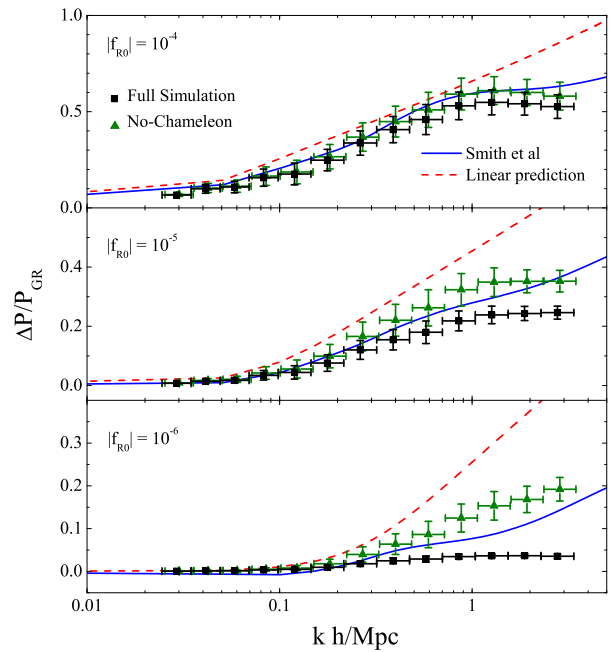}
\caption{The relative difference of matter power spectra in $f(R)$ models with respect to that in GR at redshift $z=0$. From upside down, we show the ratio $\Delta{P}/P_{\rm GR}$ for both full $f(R)$ (square) and for the non-Chameleon simulations (triangle) with $|f_{R0}|=10^{-4},10^{-5},10^{-6}$. For comparison, we over-plot the prediction from Smith et al (blue solid) and from the linear perturbation theory (red dashed). The data points with error bars are assembled from results of different box sizes based on the volume-weighted average method with a conservative $k$ cut-off. See text for details.}\label{fig:pkvw}
\end{figure}

After confirming that GR simulations are consistent with the Halofit predictions, we switch to the $f(R)$ simulations. We did the full $f(R)$, Non-Chameleon and GR simulations in three different boxes with the size $B=256,128,64$ Mpc/h, and assemble them to extend the range of scale. Generally, the $P(k)$ result is reliable up to the scale of $k_{\ast}$, which is defined as,
\be
\label{eq:kcutoff}
k_{\ast}=N_{\rm eff}{\times}k_{N/2},~~k_{N/2}\equiv\pi{N_{p}^{1/3}}/(4B),
\ee
where $N_{\rm eff}$ is a factor determined by the adaptive nature of the code, and $N_{\rm eff}=1$ for the non-adaptive simulations, \ie, the pure particle-mesh (PM) simulations. The quantity $k_{N/2}$ is the half Nyquist scale, and $N_{p}$ and $B$ are the number of particles and the box size respectively (see Table \ref{tab:nbody} for the values we use in this work). Since for the first step we aim to reproduce the results in Ref. \cite{Schmidt:2008tn} up to their resolution (half Nyquist scale), namely, $k\lesssim3$ h/Mpc, for a cross-check, we use the same $k$-cutoff criterion as Ref. \cite{Schmidt:2008tn}, namely, here we set $N_{\rm eff}=1$. We will present our results on smaller scales, which is still reliable, in the next section. Given the spectra for different boxes, we take a volume-weighted average on the overlapped scales. In Fig. \ref{fig:pkvw}, we show the matter power spectra for both the full $f(R)$ simulations and for the non-Chameleon cases at redshift $z=0$. We took a ratio with respect to the GR power spectra with the same initial condition, and then averaged over ten different realisations to reduce the sample variance, as in Refs. \cite{Oyaizu:2008sr,Oyaizu:2008tb,Schmidt:2008tn}.

As one can see, the growth is generally enhanced in $f(R)$ gravity, and the peak value of the enhancement increases as $|f_{R0}|$. This is natural bacause, for large $|f_{R0}|$, \eg, $|f_{R0}|=10^{-4}$, the Chameleon does not work, therefore the effective Newton's constant is enhanced by $4/3$ below the Compton wavelength, resulting in a large enhancement of $\Delta{P}/P_{\rm GR}$. For small $|f_{R0}|$, \eg, $|f_{R0}|=10^{-6}$, Chameleon works very efficiently, which suppresses the structure growth back to that in GR locally, yielding a tiny enhancement of $\Delta{P}/P_{\rm GR}$. The case for $|f_{R0}|=10^{-5}$ is critical -- it is something between these two extreme cases, in which Chameleon works, but not as efficient as the $|f_{R0}|=10^{-6}$ case. Therefore the difference between full and the non-Chameleon simulations in this case is intermediate.

In Fig. \ref{fig:pkvw}, we overplotted the linear prediction, and the Halofit for $f(R)$ gravity. The Halofit result is obtained by applying the Halofit fitting formula naively on the linear prediction for $f(R)$. Our simulations are consistent with
linear predictions on large scales. The Halofit prediction overestimates the deviation from GR for smaller $|f_{R0}|$.
This is because the Halofit uses only the information of linear power spectrum and fails to capture the effect of
the Chameleon mechanism, which suppresses the deviation from GR. Note also the Halofit is calibrated in GR simulations.
Thus once the deviation from GR becomes larger on smaller scales, we cannot trust the validity of the Halofit prediction.
Our result is in prefect agreement with Refs.~\cite{Oyaizu:2008tb,Schmidt:2008tn}.

\subsubsection{High-resolution results for $f(R)$}

\begin{figure*}
\includegraphics[scale=0.23]{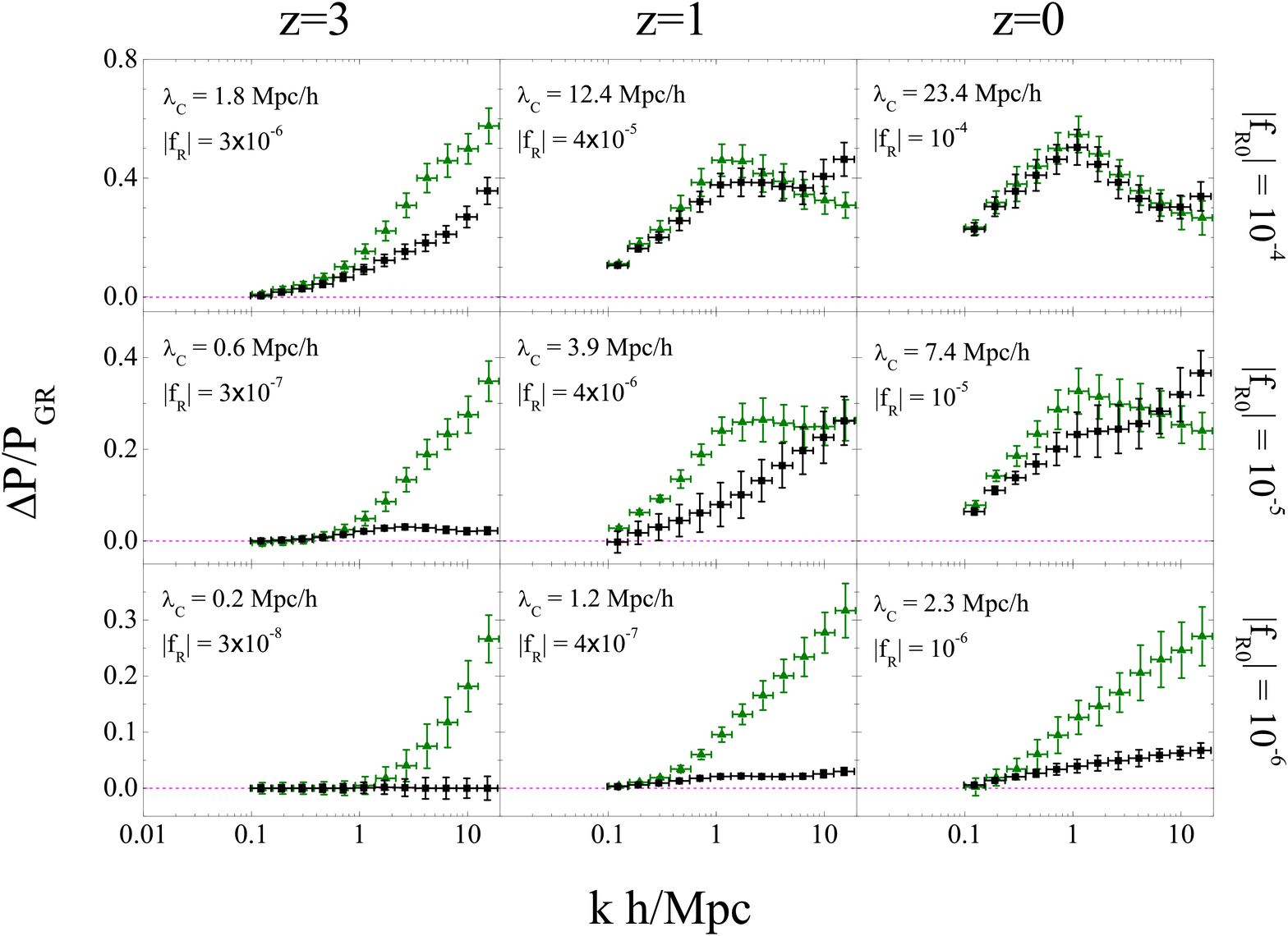}
\caption{The relative difference of matter power spectra in $f(R)$ models with respect to that in GR at redshift $z=3$ (left panels), $z=1$ (middle panel) and $z=0$ (right panels) taken from $B=64$ h/Mpc simulations. From upside down, we show the ratio $\Delta{P}/P_{\rm GR}$ for both full $f(R)$ (black squares) and for the non-Chameleon simulations (green triangles) with $|f_{R0}|=10^{-4},10^{-5},10^{-6}$. In each
panel, we show the Compton wavelength $\lambda_{\rm C}$ and the absolute value of the background scalar field $|f_R|$}
\label{fig:pk-64}
\end{figure*}

\begin{figure}
\includegraphics[scale=0.35]{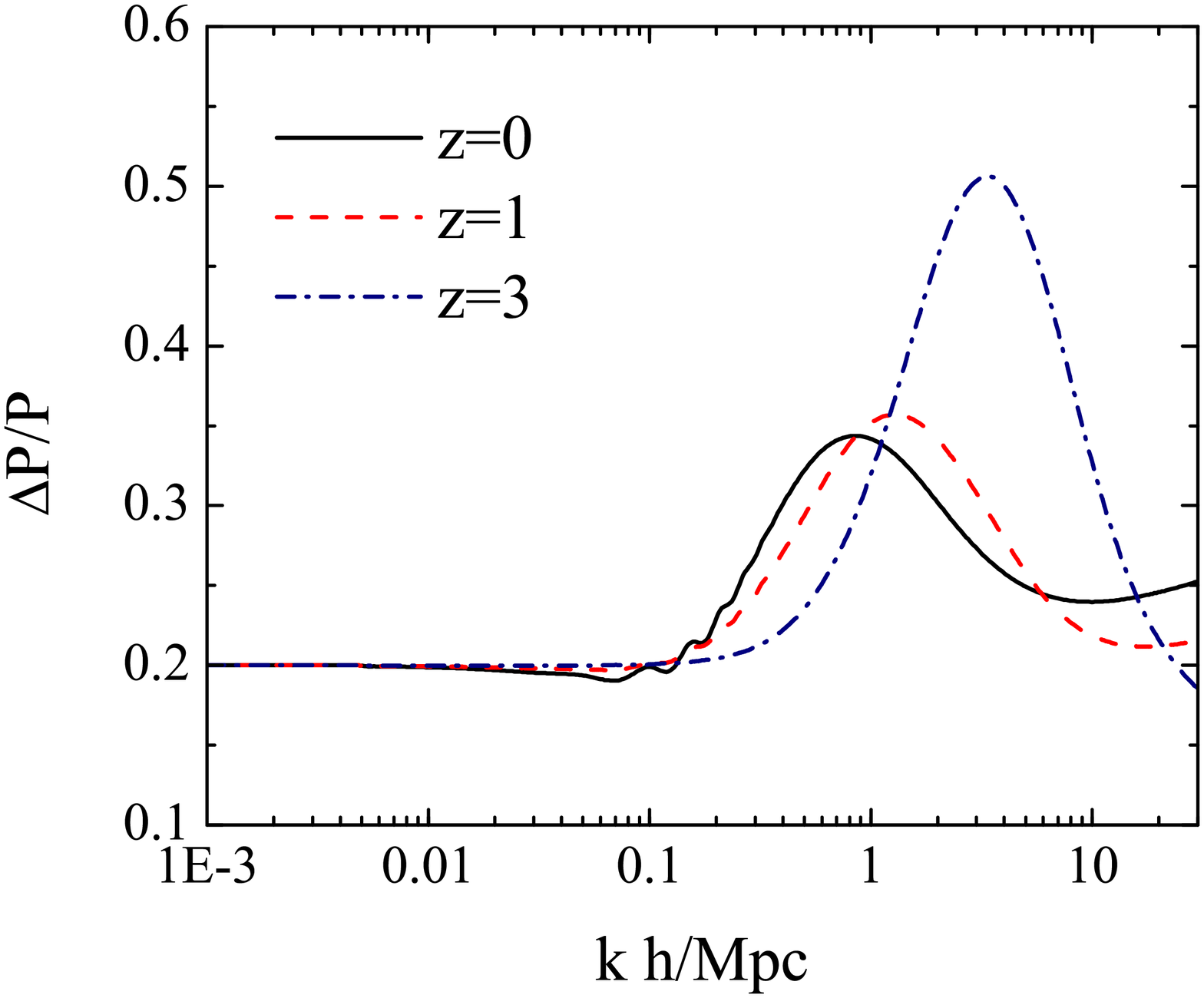}
\caption{The relative difference of matter power spectra in GR with $\sigma_8=0.9$ with respect to that with $\sigma_8=0.8$ at redshifts $z=3$ (blue dash-dot), $z=1$ (red dash) and $z=0$ (black solid). Halofit is used to calculate for the nonlinearities.  }\label{fig:pk-norm}
\end{figure}

The self-adaptive nature of our code allows us to go beyond the half Nyquist scale, in other words, $N_{\rm eff}$ can be greater than $1$. The realistic value of $N_{\rm eff}$ is largely determined by the maximum number of refinement triggered in the whole simulation process, which is $8\sim10$ for our simulations. To be conservative, we choose $N_{\rm eff}=7$ in all cases, and show the high-resolution result for power spectra in Figs. \ref{fig:pk-64}. As in Fig. \ref{fig:pkvw}, we show the relative difference of power spectra in $f(R)$ with respect to that in GR at various redshifts.

Let us see the higher resolution result from $B=64$ Mpc/h simulations shown in Fig. \ref{fig:pk-64} at $z=3, 1$ and $0$.
Looking at the non-Chameleon result first, one could have the following observations,

  \begin{enumerate}
    \item The ratio increases as one goes to small scales, and it goes down after reaching a peak at $z=0$ and $z=1$ for N4 and N5 cases;
    \item For the same model, the peak moves towards larger $k$ as one goes to higher redshift;
    \item The ratio monotonically increases with $k$ at $z=3$ for N4 and N5 cases and at all times for N6 cases.
  \end{enumerate}

One could use GR as an example to understand these observations. It is true that GR is different from our non-Chameleon $f(R)$ models, but the physical origin of these features are similar. In Fig. \ref{fig:pk-norm}, we show the relative difference of matter power spectra in GR with $\sigma_8=0.9$ with respect to that with $\sigma_8=0.8$ at redshifts $z=3,1$ and $0$. We use {\tt CAMB} \footnote{Available at \url{http://camb.info}} to generate the linear power spectra, and use Halofit to account for the nonlinearities.

Interestingly, the pattern here is similar to what we saw just now -- the ratio in GR shows a peak, and as $z$ increases, the peak becomes more pronounced, and it shifts to larger $k$. This is due to the transition between the 2-halo term to the 1-halo term in the halo model description of the power spectrum (for a review see \cite{Cooray:2002dia}). If $\sigma_8$ is higher, the non-linearity becomes important earlier and the transition from 1 to 2-halo term appears at smaller $k$. Since the 2-halo term has a larger amplitude, this creates a peak in the ratio. As the transition wave number between 1 and 2-halo terms moves to smaller $k$ for smaller $z$, the peak is shifted to smaller $k$. The same thing happens in non-Chameleon simulations. In this case, the linear power has the $k$-dependent enhancement. The transition from 1-halo to 2-halo term appears at high $k$ at $z=3$ in $f(R)$ models but we do not see the peak as the $k$-dependent enhancement of the linear power spectrum hides the enhancement due the 2-halo term. At $z=0$, the transition appears at smaller $k$ where the linear enhancement is weak then we see the peak due to the transition from 1 to
2-halo term.

Having understood the non-Chameleon simulations results, now let us look at the full Chameleon simulations. At $z=3$, we can
see that the Chameleon mechanism works well even in the F4 case, and the deviation from GR is effectively suppressed. Once the Chameleon does not work well, \eg, at $z=0$ for F4, the full simulation result traces the non-Chameleon result as expected, with slightly lowered amplitude. Interestingly when the Chameleon mechanism has started to fail, {\it e.g.} at $z=1$ for F4 and at $z=0$ for F5, the power spectrum in full simulations has larger amplitudes than those in the non-Chameleon simulations on small scales. This implies that complicated dynamics is taking place when the Chameleon mechanism fails to work. We will come back
to this issue later by studying the density profiles of halos. For the case where the Chameleon mechanism works very well, \ie, the F6 case, the full simulation result has much less amplitudes, implying that GR is successfully recovered.

\subsubsection{PPF fit}

\begin{figure*}
\includegraphics[scale=0.2]{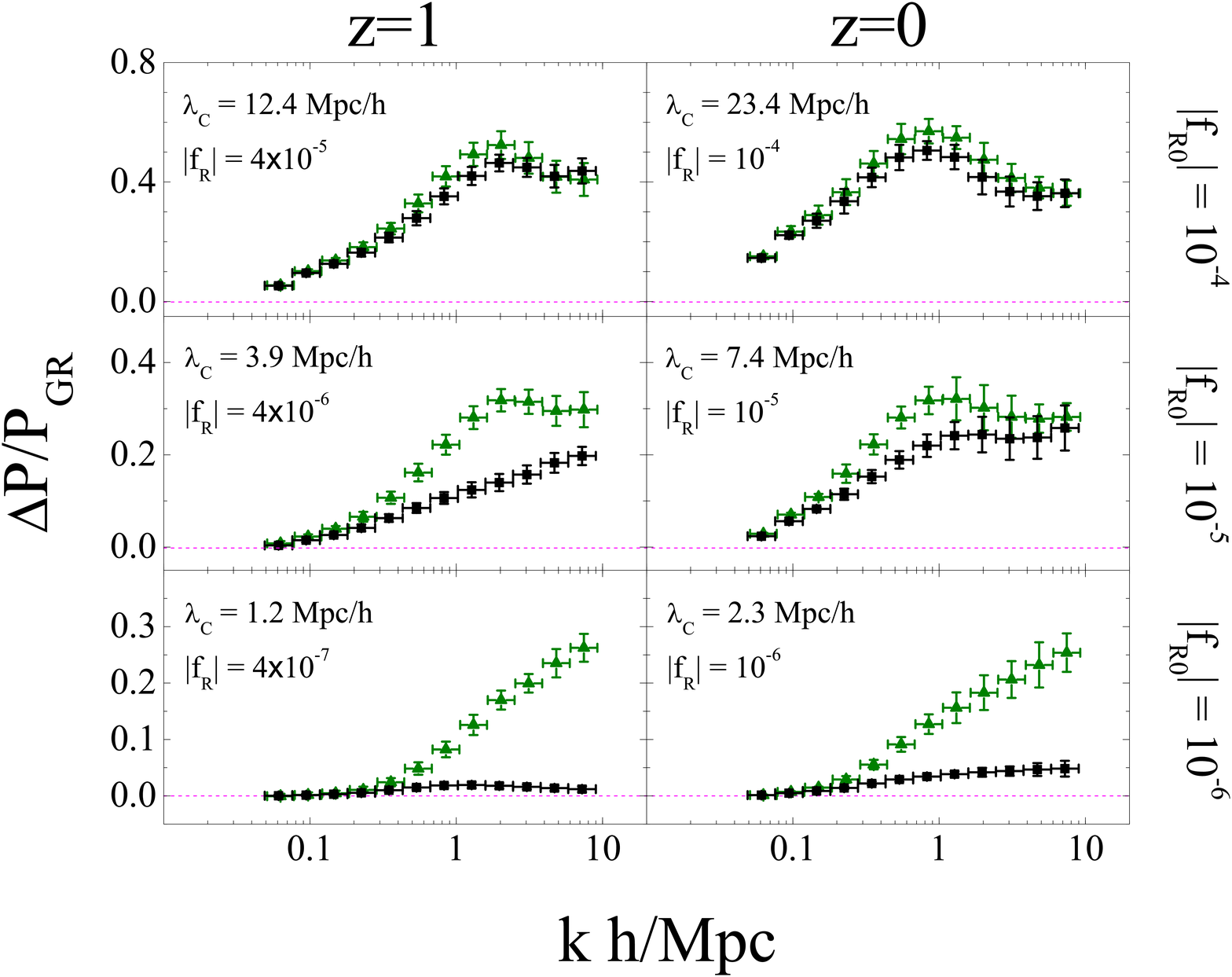}
\caption{Same as Fig. \ref{fig:pk-64} but for $B=128$ h/Mpc simulations at redshifts $z=1$ and $0$.}
\label{fig:pk-128}
\end{figure*}

\begin{figure*}
\includegraphics[scale=0.25]{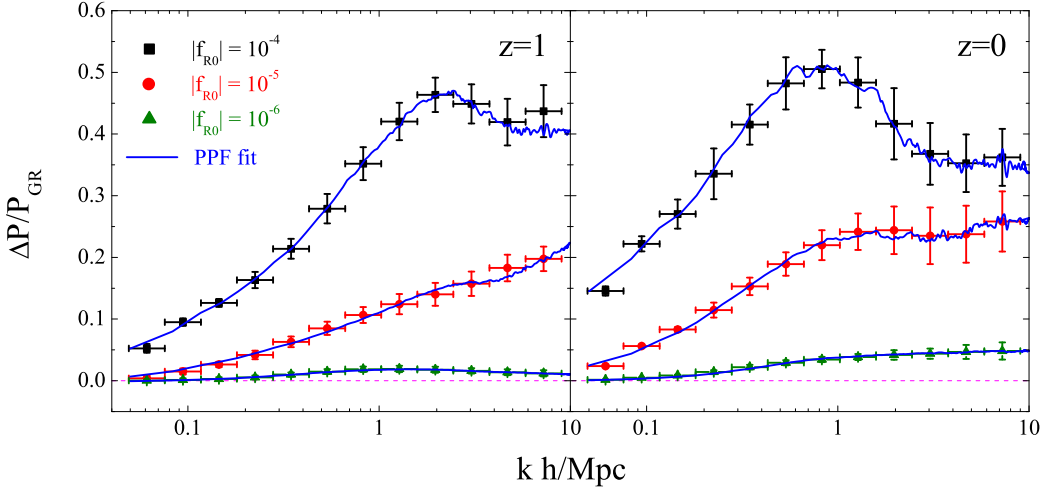}
\caption{The relative difference of matter power spectra in full $f(R)$ simulations with respect to $P(k)$ in GR at redshift $z=1$ (left panel) and $z=0$ (right panel). The results for three $f(R)$ models $|f_{R0}|=10^{-4},10^{-5},10^{-6}$ are shown in black blocks, red circles and green triangles respectively. The blue solid curves show the PPF fit, and the horizontal dashed line illustrates $\Delta{P}=0$.}\label{fig:pkPPF}
\end{figure*}

Hu and Sawicki proposed a fitting formula (PPF fit \cite{Hu07}) for the matter power spectrum in modified gravity
based on the assumption that the power spectrum resembles that in GR on small scales, which takes the form of,
\be\label{eq:ppf}
P(k,z)=\frac{P_{\rm non-GR}(k,z)+c_{\rm nl}\Sigma^2(k,z)P_{\rm GR}(k,z)}{1+c_{\rm nl}\Sigma^2(k,z)},
\ee
where $P_{\rm GR}$ denotes the non-linear power spectrum in a $\Lambda$CDM model with the same expansion history as that in the modified gravity models, and $P_{\rm non-GR}$ means the nonlinear power spectrum in modified gravity {\it without} the mechanism that recovers GR on small scales, {\it i.e.} the Chameleon mechanism.
 Therefore in our case, $P_{\rm non-GR}$ is exactly the power spectra for the non-Chameleon simulation, \ie, $P_{\rm non-GR}=P_{\rm no-cham}$. $\Sigma$ measures the non-linearities and once the non-linearities are significant
$\Sigma \gg 1$, the power spectrum approaches $P_{GR}$.
Koyama \etal~\cite{Koyama:2009me} did a successful PPF fit and recovered the full $f(R)$ simulation up to $k\sim0.5$ h/Mpc presented in \cite{Oyaizu:2008tb,Schmidt:2008tn} using Eq. (\ref{eq:ppf}) with $\Sigma$ given by
\be
\label{eq:Sigma}
\Sigma^2(k,z)=\left[\frac{k^3}{2\pi^2}P_{\rm lin}(k,z)\right]^{1/3},
\ee
where $P_{\rm lin}$ is the linear power spectrum in $f(R)$.

Since our simulation goes to much smaller scales than \cite{Oyaizu:2008tb,Schmidt:2008tn}, we study whether we can extend this fitting formula to smaller scales. To be conservative, we use the results from $B=128$ Mpc/h simulations to extend the formula
up to $k=10$ h/Mpc (see Fig.~\ref{fig:pk-128}). We generalise Eq. (\ref{eq:Sigma}) by adding three more parameters
$\alpha, \beta$ and $\gamma$,
\be
\label{eq:Sigma-gen}
\Sigma^2(k,z)=\left[\frac{k^3}{2\pi^2}P_{\rm lin}(k,z)\right]^{\alpha+\beta{k}^\gamma}.
\ee
Combining Eqs. (\ref{eq:ppf}) with (\ref{eq:Sigma-gen}),
we fit $\Delta P/P_{\rm GR}$ up to $k\sim10$ h/Mpc, and the best-fit parameters are listed in Table \ref{tab:ppf}. The best-fit power spectrum curves for $f(R)$ models are overplotted with the simulation data in Fig.
\ref{fig:pkPPF}. As one can see, the agreement for F6 model is within one percent level even if we fixed $\beta$ and $\gamma$ to zero. This implies that, if the Chameleon works, we can use the PPF formulae by varying $c_{\rm nl}$ and $\alpha$ only. However, if the Chameleon fails, the power spectrum tends to go back to the non-Chameleon spectrum
on small scales, so that we have to introduce additional parameters $\beta$ and $\gamma$ to parametrise this behaviour.
Once we include these parameters, we can fit the power spectrum in F4 and F5 models accurately. In the quasi non-linear
regime $k<1$ h/Mpc, we can ignore the corrections described by $\beta$ and our results are roughly consistent with \cite{Koyama:2009me}. The PPF fitting formula would be useful when we try to find fitting fomulae similar to Halofit
in $f(R)$ gravity models because the non-Chameleon simulations are far easier to run because of the linearity of the equation and the cosmological parameter dependence of the PPF parameters was found to be weak at least in the quasi non-linear regime.

\begin{table}
\begin{tabular}{c|ccc|ccc}
  \hline  \hline
 Redshift  & \multicolumn{3}{c|}{$z=1$} & \multicolumn{3}{c}{$z=0$} \\

  $|f_{R0}|$& $10^{-4}$ &$10^{-5}$ & $10^{-6}$ & $10^{-4}$ &$10^{-5}$ & $10^{-6}$ \\
  \hline
  $c_{\rm nl}$& $0.17$ &$1.1$ & $3.4$ & $0.09$ &$0.44$ & $1.6$ \\
  $\alpha$& $0.42$ &$2.5$ & $1.3$ & $0.26$ &$0.76$ & $0.39$ \\
  $\beta$& $-0.22$ &$-2.2$ & fixed to $0$ & $-0.01$ &$-0.85$ & fixed to $0$ \\
  $\gamma$& $1.6$ &$0.15$ & fixed to $0$ & $2.04$ &$0.24$ & fixed to $0$ \\

  \hline  \hline
\end{tabular}
\caption{The best-fit PPF parameters for various $f(R)$ models at redshifts $z=1$ and $z=0$.}\label{tab:ppf}
\end{table}

\subsection{Mass Function}

\subsubsection{GR}

\begin{figure}[t]
\includegraphics[scale=0.2]{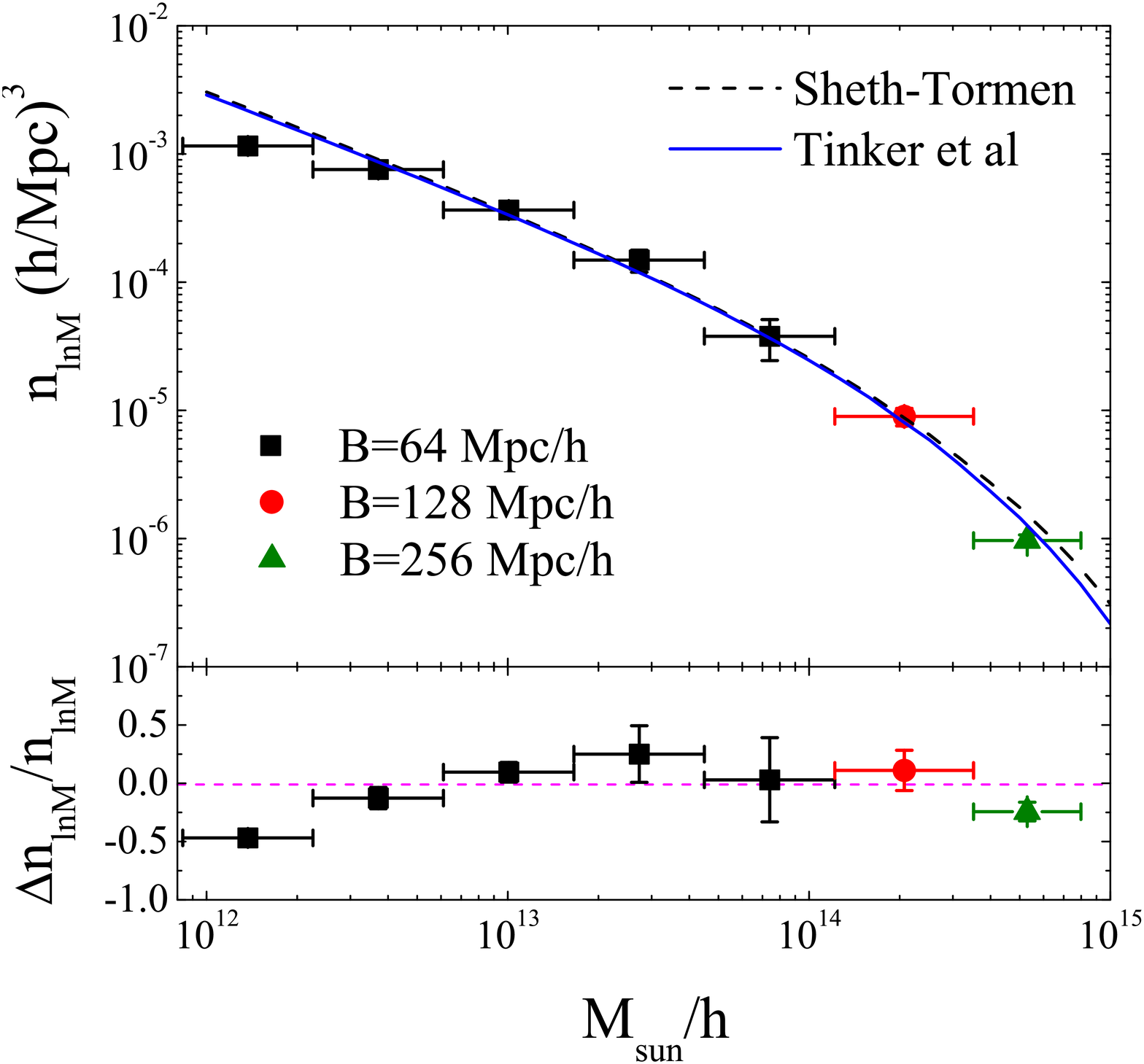}
\caption{Upper panel: the mass function for GR at redshift $z=0$ stacked using $B=64$ Mpc/h (black blocks), $128$ Mpc/h (red circle) and $256$ Mpc/h (green triangle) overplotted with the theoretical predications from Sheth-Tormen (black dash) and from Tinker \emph{et al.} (blue solid); Lower panel: relative difference of the simulation result with respect to the Tinker \emph{et al.} prediction.}\label{fig:GRmf}
\end{figure}

We test our mass function in GR firstly by comparing it to the theoretical predictions proposed by Sheth and Tormen \cite{Sheth:1999mn}, and by Tinker \etal \cite{Tinker:2008ff}. The comparison is shown in Fig. \ref{fig:GRmf}. We stack our results from different boxes to extend the mass range, and we see an agreement with theoretical predictions. It seems that our simulation underestimates the number of small halos (halos in the first bin) because of the lack of particles, but this problem can be alleviated to some extent when taking the ratio of the mass function of $f(R)$ with respect to that in GR.

\subsubsection{$f(R)$}

\begin{figure*}[t]
\includegraphics[scale=0.3]{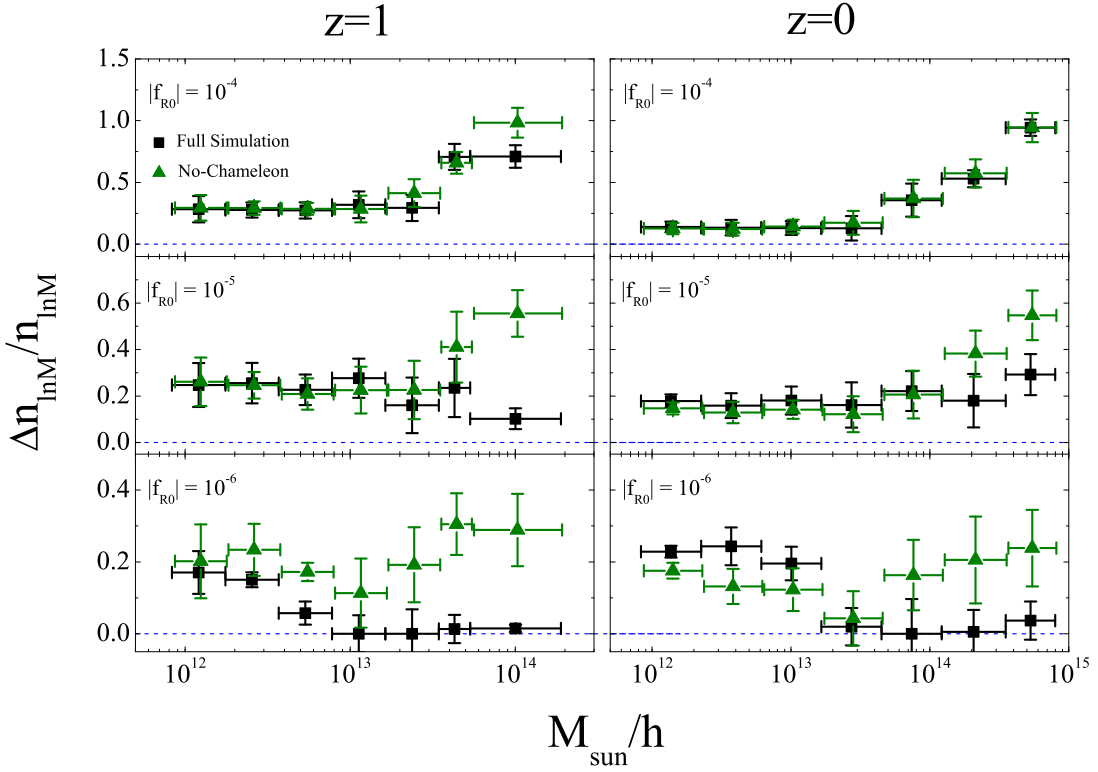}
\caption{The relative difference of the mass function for $f(R)$ models with respect to that in GR at redshift $z=1$ (left) and $z=0$ (right). The results for the full $f(R)$ simulations (black blocks) and non-Chameleon cases (green triangles) of $|f_{R0}|=10^{-4},10^{-5},10^{-6}$ are shown from upside down. The results for small halos (first three mass bins from the left), intermediate-sized halos (three bins in the middle), and the large halos (three bins on the right) are obtained from the simulations with box size $B=64, 128$ and $256$ Mpc/h respectively.}\label{fig:mf}
\end{figure*}

The result for the halo mass function is presented in Fig. \ref{fig:mf}. As one can see, our result at $z=0$ is consistent with that in Ref. \cite{Schmidt:2008tn} (cf Fig. 2). Let us understand our result by looking at the non-Chameleon result first. In this case, the strength of gravity is enhanced by $4/3$ below the Compton wavelength $\lambda_c$, regardless of the value of $|f_{R0}|$. Meanwhile, $\lambda_c$ increases with $|f_{R0}|$, making the enhancement on the mass function more pronounced for large $|f_{R0}|$ models. This is what we see in Fig.~\ref{fig:mf}. Also, at redshift $z=0$, the enhancement for larger halos is more significant, while at redshift $z=1$, smaller halos are more populated. This is because the stronger gravity in $f(R)$ models create more small halos at high redshift and assemble them to make more large halos at low redshift compared with $\Lambda$CDM models.

For the full $f(R)$ simulations at $z=0$, the number of massive halos is generally reduced compared to the non-Chameleon model predictions, because the suppression of the fifth force due to the Chameleon mechanism is stronger in high density regions, where massive halos most likely reside. For the case of $|f_{R0}|=10^{-4}$, in which Chameleon does not work, the full and non-Chameleon results agree well, while for the $|f_{R0}|=10^{-6}$ case, the number of large halos is almost the same as the GR prediction due to the Chameleon effect. At redshift $z=1$, the suppression starts from lower mass bins simply because Chameleon was working more efficiently at higher redshift.

\subsection{Halo Profile}

\begin{figure*}
\includegraphics[scale=0.25]{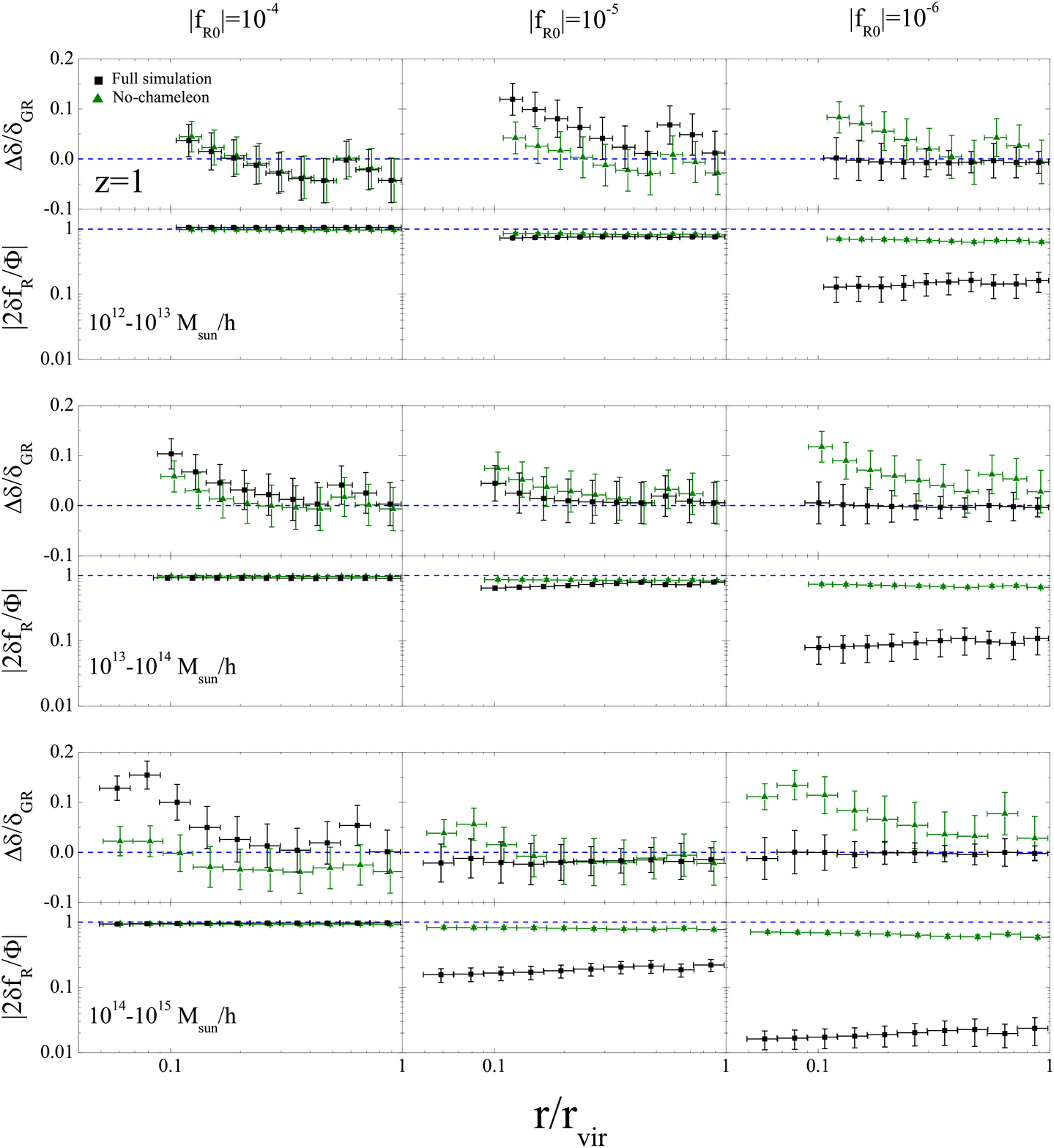}
\caption{The profiles of density and $|2\delta{f_R}/\Phi|$ as a function of $r/r_{\rm vir}$ for full $f(R)$ simulations (black blocks) and non-Chameleon cases (green triangles) for models with $|f_{R0}|=10^{-4},10^{-5},10^{-6}$ (from left to right) at redshift $z=1$. For the density profiles, we show the relative difference with respect to that in GR, as we did for matter power spectra and mass functions.}\label{fig:prof-z1}
\end{figure*}

\begin{figure*}
\includegraphics[scale=0.25]{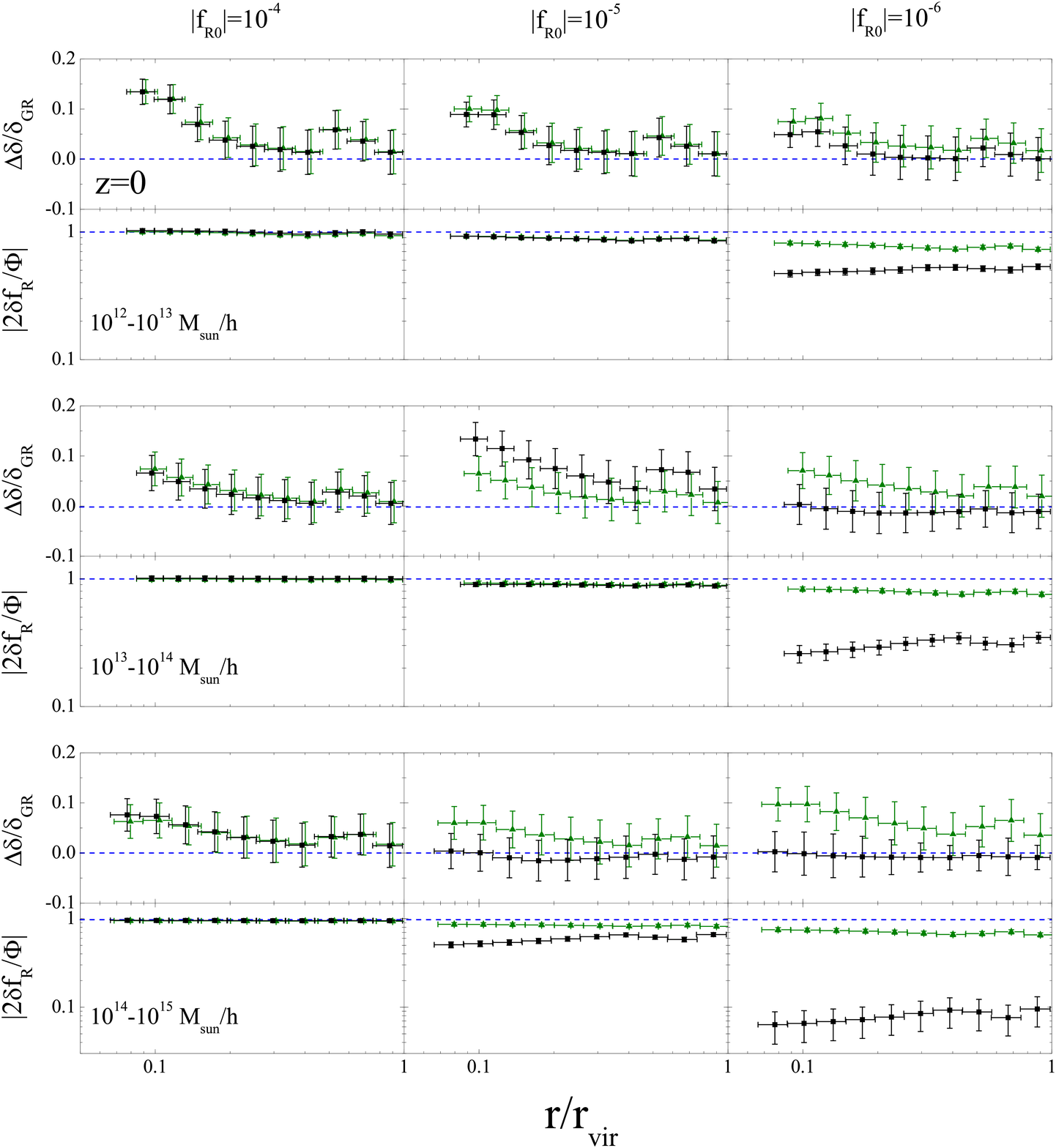}
\caption{Same as Fig. \ref{fig:prof-z1}, but at redshift $z=0$.}\label{fig:prof}
\end{figure*}

We show the halo profile for the overdensity and for $\Gamma$ at redshift $z=1$ and $z=0$ in Figs.~\ref{fig:prof-z1} and \ref{fig:prof} respectively. We show the results for the halos in three mass ranges: $[10^{12},10^{13})$, $[10^{13},10^{14})$ and $[10^{14},10^{15}] M_{\odot}$/h, taken from the $B=64,128$ and $256$ simulation boxes respectively.

A quick observation of the result is that on average, the profiles of the overdensity for $f(R)$ gravity are similar to that in GR, namely, the relative difference is smaller than 20\% in all cases. This is consistent with the analysis in Ref.~\cite{Schmidt:2008tn}. The complex patterns of the relative difference of $f(R)$ and GR halo density profiles can be understood qualitatively as follows.
\begin{enumerate}
\item For the F6 simulation, the Chameleon effect is very efficient throughout the cosmic history and so the fifth force has always been strongly suppressed, so that the predicted halo density profile should be very close to the GR result, which is what we see from Figs.~\ref{fig:prof-z1} and \ref{fig:prof} (the only exception is the case of low mass halos at $z=0$, which is because these halos have lower density and so the Chameleon effect is weaker, especially at late times).
\item For the N6 simulation, the fifth force is only weakly suppressed (especially at late times), and the central attractive potential towards the halo centre is stronger than in GR, which means that particles tend to fall towards the halo centres, producing higher density profiles than the latter.
\item The N5 and N4 results follow the same pattern as N6, but are more complicated as in some cases the density profile can be even lower than in GR. This is perhaps because the central attractive potential is not the only thing affecting the halo density profile. As the fifth force is unsuppressed from early times, the particles are typically faster than they are in GR, and  tend to escape the halo, flattening the density profile. Note that fifth force both deepens the central potential and speeds up the particles, and the latter is an accumulative effect, which makes it difficult to see which effect is dominating.
\item The F4 simulation agrees with N4 simulation very well at late time, because Chameleon effect is insignificant. But at early times it produces higher density profile than N4, except for the small halos (for which Chameleon effect is again unimportant), which seems to indicate that the particles in the F4 halos are still relatively slow, because the fifth force has not become strong enough for long.
\item The case of F5 is further complicated by the fact that $|f_{R0}|=10^{-5}$ is a critical point for our $f(R)$ models. For the small halos, the fifth force has just become unsuppressed at $z=1$, and we see that their density profiles are higher than N5 results for the same reason as above, while at $z=0$ the F5 and N5 simulations agree very well. For the medium-sized halos, at $z=1$ the fifth force is still strongly suppressed and their density profiles are lower than N5 results because the central attractive potential is weaker, while at $z=0$ the fifth force becomes unsuppressed and the density profiles are higher than N5 results again for the same reason as above. For the most massive halos, the fifth force is strongly suppressed even at present, and the halo density profiles are lower than in N5 for both $z=1$ and $z=0$.
\end{enumerate}
Therefore we see the following general evolution pattern when comparing the halo density profiles from F and N simulations: at early times the fifth force is strongly suppressed in F simulations, and so the central potential is weaker there, producing lower density profiles; then the fifth force becomes unsuppressed and the central potential becomes as strong as in N simulations, while the particles are still moving relatively slowly, producing higher density profiles. This happens at $z=1$ in F4 simulations and $z=0$ in F5 simulations. Finally the particles move fast due to the fifth force, and the density profiles approach the non-Chameleon results as is seen in F4 simulations at $z=0$. The transition happens earlier for models with larger $|f_{R0}|$ and for smaller halos (in both cases the fifth force becomes unsuppressed earlier). What we see in Figs.~\ref{fig:prof-z1} and \ref{fig:prof} are just different stages of the above evolution pattern.

This picture is consistent with the behaviour of the power spectrum shown in Fig \ref{fig:pk-64}. Once the Chameleon mechanism fails and the density profile becomes higher than those in N simulations, the power spectrum in F simulations tend to be higher on smaller scales (at $z=1$ in F4 simulations and at $z=0$ in F5 simulations). Then it approaches non-Chameleon results later (at $z=0$ in F4 simulations). We expect to see the same evolution pattern in F5 and F6 simulations if we would continue our simulations in the future.

In Figs.~\ref{fig:prof-z1} and \ref{fig:prof} we have also shown the efficiency $\Gamma$ of the Chameleon effect. For the N4 and F4 simulations, we see that $\Gamma=1$ almost perfectly for both $z=1$ and $z=0$, indicating that the Chameleon mechanism does not work for $|f_{R0}|=10^{-4}$. Note that the agreement between N4 and F4 serves as an independent test of our code, as the treatments for these simulations are very different. For the N5 and F5 simulations, we see that $\Gamma$ is close to one for the small and medium-sized halos at both redshifts, while it is significantly less than $1$ for the massive halos, especially at $z=1$, which are all as expected because the Chameleon effect is stronger in high density regions and at early times. $\Gamma$ is not perfectly equal to $1$ even at late times and in small halos, because there {\it is} a suppression of the fifth force and $|\delta f_R|$ even in the linearised treatment above the Compton wavelength. In the N6 and F6 simulations, we see that $\Gamma\ll1$ for all halos and at all times, showing a strong Chameleon effect.

Interestingly, in the F simulations $\Gamma$ increases with the distance from the halo centre, while in N simulations the trend is just opposite (this is clearest in the F6/N6 simulations at $z=0$). The former is because the Chameleon effect gets weaker as one moves from the halo centre (highest density region) and so $\Gamma$ tends to $1$; the latter is because the value of $\delta f_R$ is only affected by particles lying within the Compton wavelength from the considered position and as one moves from the halo centre more and more particles in the central region of the halo become unable to affect $\delta f_R$, while this does not happen to $\Phi$ as gravity is a long-range force.

In summary, we find that the halo density profiles in the $f(R)$ gravity model show complicate but interesting features, which could be understood qualitatively. We would like to study these in more details in future works, perhaps with the aid of even-higher-resolution simulations.

\section{Conclusion}

Modified gravity scenario, especially the $f(R)$ gravity, attracts more and more attention as an alternative to dark energy to explain the origin of the cosmic acceleration at low redshifts. Much efforts have been made to investigate the phenomenology of $f(R)$ theories on linear scales \cite{Song:2007, Li:2007, Hu:2007nk,Hu07,Pogosian:2007sw,Zhao:2008bn}. Unfortunately, the current observational data on linear scales, including weak lensing, Integrated Sachs-Wolfe (ISW) and cosmic microwave background (CMB), \etc, can only weakly constrain $f(R)$ gravity \cite{Song:2007da,Giannantonio:2009gi,Lombriser:2010mp}. And even futuristic observations on linear scales can hardly prove, or falsify the $f(R)$ scenario \cite{Zhao:2008bn}.

On nonlinear scales, $f(R)$ gravity can be much better tested by cosmological observations \cite{Schmidt:2009am,Lombriser:2010mp} because the scale dependent enhancement of the growth rate makes deviations
from GR larger and larger on smaller scales. However, we should take into account the mechanism to recover GR that is
necessary to evade the strong constraints in the solar system. In $f(R)$ gravity models, this recovery of GR is accomplished
by the Chameleon mechanism. In order to realise the Chameleon mechanism, the scalar mode in $f(R)$ gravity should satisfy
a highly non-linear evolution equation.  In this work, we implement the Newton-Gauss-Seidel relaxation solver into the \texttt{MLAPM} code, an adaptive particle mesh simulation code, and obtain high resolution simulation results (up to $k\sim20$ h/Mpc for the matter power spectrum). We independently confirmed the results, including the power spectrum and halo statistics, presented in Refs. \cite{Oyaizu:2008tb} and \cite{Schmidt:2008tn}, and extended the resolution by a factor of $7$.

Deviations from GR in the power spectrum are highly suppressed at early times when the Chameleon mechanism works very well.
Later the Chameleon starts to fail once the background $f_R$ field becomes comparable to its fluctuations, $\delta f_R$. Then the power spectrum approaches the one in the case without the Chameleon mechanism. This transition happens
at a higher redshift for a larger $|f_{R0}|$ simply because the background field is larger for larger $|f_{R0}|$. We found an interesting behavior
that the power spectrum in full Chameleon simulations has higher amplitude than the one in non-Chameleon simulations
during this transition. We observed a similar behavior in the density profile. The density profile in full simulations
becomes higher than the one in non-Chameleon simulations on small scales once the Chameleon has started to fail.
Qualitatively, this is due to the fact that, once the Chameleon mechanism fails, the fifth force becomes unsuppressed and the central potential
of halos becomes as strong as in the non-Chameleon simulations but the particles are moving still relatively slowly.
Then dark matter particles are temporarily trapped at the central region of halos. Later particles velocities catch up and the density profile and the power spectrum approach those in non-Chameleon simulations. In order to confirm this picture, we would need higher resolution simulations and study the behavior of dark matter particles in halos carefully. We can also make the connection between the halo density profile and the power spectrum using the halo model approach. We leave the study of these issues in a separate paper.

We measured the profile of the scalaron field inside halos and studied the efficiency of the Chameleon mechanism
As expected, we found that the Chameleon mechanism works better for heavier halos because the density is high in these
halos. The halo mass function also shows the same tendency -- the number of heavier halos approaches that in GR simulations
if the Chameleon mechanism works as in the $|f_{R0}|=10^{-6}$ case. For the $|f_{R0}|=10^{-4}$ case,
the Chameleon no longer works at $z=0$, which results in more and more heavier halos compared with GR.

In conclusion, we studied the effect of the Chameleon mechanisms in details in our high resolution simulations.
It was found that once the Chameleon mechanism
stated to fail, the power spectrum and halo properties showed very interesting behaviours before they approach those
in non-Chameleon simulations. For the power spectrum, we showed that this transition can be described by extending
the Post-Parametrised-Friedmann (PPF) formalism that interpolates between the power spectrum in non-Chameleon models
and GR models. This kind of fitting formulae will be useful when we confront our predictions to observations
as it is far easier to run non-Chameleon simulations with different cosmological parameters.
Our results indicate that $|f_{R0}|=10^{-5}$ is the critical case where the Chameleon mechanism fails to
work today. It becomes significantly harder to detect deviations from GR once $|f_{R0}|$ becomes smaller
than $10^{-5}$. In this case, we need to look into the places where the Chameleon still fails to work, {\it i.e.}
smaller halos. We leave a study of observational constraints on $f(R)$ gravity using non-linear clustering such
as weak lensing and cluster and voids abundance in a forthcoming paper.

\acknowledgements

We thank Fabian Schmidt for enlightening discussions. GBZ thanks DAMTP of University of Cambridge for hospitality. GBZ and KK are supported by STFC grant ST/H002774/1. KK also acknowledges support from the European Research Council, the Leverhulme trust and RCUK. BL is supported by Queens' College at University of Cambridge and STFC.

\appendix

\section{Discretisation of Equations}

\label{appen:discrete}

The last thing we are going to do before implementing the $f(R)$ equations into the $N$-body code is to discretise them, such that they fit to the philosophy of the numerical computations -- solving the equations on meshes with finite grid sizes. This appendix displays the discretised equations, and again we discuss the Chameleon and non-Chameleon cases separately. We shall only write down the discrete equations for $u$ or $\dfR$, as those for $\Phi_c$ as simple.

\subsection{Full $f(R)$ simulations}

The full equation for $u$, Eq.~(\ref{eq:sf_codeunit_cham}), contains the quantity $\nabla^2e^u=\nabla\cdot\left(e^u\nabla u\right)$. To discretise it, let us define $b\equiv e^u$, and assumed that the discretisation is performed on a grid with grid spacing $h$. We shall require second order precision which is the same as the default Poisson solver in {\tt MLAPM}, and then $\nabla u$ in one dimension can be written as
\be
\nabla u\ \rightarrow\ \nabla^h u_j\ =\ \frac{u_{j+1}-u_{j-1}}{2h},
\ee
where a subscript $j$ means that the quantity is evaluated on the $j$-th point. The generalisation to three dimensional case is straightforward.

The factor $b$ in $\nabla\cdot(b\nabla u)$ makes this a standard variable coefficient problem. We need also discretise $b$, and do it in this way (again for one dimension) \cite{Li:2009sy}:
\begin{eqnarray}
&&\nabla\cdot(\nabla u)\nonumber\\
&\rightarrow& \frac{1}{h^2}\left[b_{j+\frac{1}{2}}u_{j+1} - u_j\left(b_{j+\frac{1}{2}}+b_{j-\frac{1}{2}}\right) + b_{j-\frac{1}{2}}u_{j-1}\right], \ \ \
\end{eqnarray}
in which $b_{j\pm\frac{1}{2}}=\frac{1}{2}\left(b_j+b_{j\pm1}\right)$. Generalising this to three dimensions, we have
\begin{widetext}
\begin{eqnarray}
\nabla\cdot(\nabla u)
&\rightarrow& \frac{1}{h^2}\left[b_{i+\frac{1}{2},j,k}u_{i+1,j,k} - u_{i,j,k}\left(b_{i+\frac{1}{2},j,k}+b_{i-\frac{1}{2},j,k}\right) + b_{i-\frac{1}{2},j,k}u_{i-1,j,k}\right]\nonumber\\
&&+\frac{1}{h^2}\left[b_{i,j+\frac{1}{2},k}u_{i,j+1,k} - u_{i,j,k}\left(b_{i,j+\frac{1}{2},k}+b_{i,j-\frac{1}{2},k}\right) + b_{i,j-\frac{1}{2},k}u_{i,j-1,k}\right]\nonumber\\
&&+\frac{1}{h^2}\left[b_{i,j,k+\frac{1}{2}}u_{i,j,k+1} - u_{i,j,k}\left(b_{i,j,k+\frac{1}{2}}+b_{i,j,k-\frac{1}{2}}\right) + b_{i,j,k-\frac{1}{2}}u_{i,j,k-1}\right].
\end{eqnarray}
\end{widetext}
Then the discrete version of Eq.~(\ref{eq:sf_codeunit_cham}) is
\begin{eqnarray}\label{eq:diffop}
L^{h}\left(u_{i,j,k}\right) &=& 0,
\end{eqnarray}
in which
\begin{widetext}
\begin{eqnarray}
L^{h}\left(u_{i,j,k}\right) &=&
\frac{1}{h^{2}}\frac{ac^2}{(BH_0)^2}\left[b_{i+\frac{1}{2},j,k}u_{i+1,j,k}
- u_{i,j,k}\left(b_{i+\frac{1}{2},j,k}+b_{i-\frac{1}{2},j,k}\right)
+ b_{i-\frac{1}{2},j,k}u_{i-1,j,k}\right]\nonumber\\
&& +
\frac{1}{h^{2}}\frac{ac^2}{(BH_0)^2}\left[b_{i,j+\frac{1}{2},k}u_{i,j+1,k}
- u_{i,j,k}\left(b_{i,j+\frac{1}{2},k}+b_{i,j-\frac{1}{2},k}\right)
+ b_{i,j-\frac{1}{2},k}u_{i,j-1,k}\right]\nonumber\\
&& +
\frac{1}{h^{2}}\frac{ac^2}{(BH_0)^2}\left[b_{i,j,k+\frac{1}{2}}u_{i,j,k+1}
- u_{i,j,k}\left(b_{i,j,k+\frac{1}{2}}+b_{i,j,k-\frac{1}{2}}\right)
+ b_{i,j,k-\frac{1}{2}}u_{i,j,k-1}\right]\nonumber\\
&&-\Omega_{m}\rho_{c,i,j,k}-4a^3\Omega_{\Lambda}+\frac{1}{3}\Omega_{m}a^3
\Big(n\frac{c_1}{c_2^2}\Big)^{\frac{1}{n+1}}e^{-\frac{u_{i,j,k}}{n+1}}.
\end{eqnarray}
\end{widetext}
Then the Newton-Gauss-Seidel iteration says that we can obtain a new (and usually more accurate) solution of $u$, $u^{\rm new}_{i,j,k}$, using our knowledge about the old (and less acurate) solution $u^{\rm old}_{i,j,k}$ as
\begin{eqnarray}\label{eq:GS}
u^{\mathrm{new}}_{i,j,k} &=& u^{\mathrm{old}}_{i,j,k} -
\frac{L^{h}\left(u^{\mathrm{old}}_{i,j,k}\right)}{\partial
L^{h}\left(u^{\mathrm{old}}_{i,j,k}\right)/\partial u_{i,j,k}}.
\end{eqnarray}
The old solution will be replaced with the new one once the latter is ready, using a red-black Gauss-Seidel sweeping scheme. Note that
\begin{widetext}
\begin{eqnarray}
\frac{\partial L^{h}(u_{i,j,k})}{\partial u_{i,j,k}} &=&
\frac{1}{2h^{2}}e^{u_{i,j,k}}\frac{ac^2}{(BH_0)^2}\left[u_{i+1,j,k}+u_{i-1,j,k}+u_{i,j+1,k}
+u_{i,j-1,k}+u_{i,j,k+1}+u_{i,j,k-1}-6u_{i,j,k}\right]\nonumber\\
&&-\frac{1}{2h^{2}}\frac{ac^2}{(BH_0)^2}\left[b_{i+1,j,k}+b_{i-1,j,k}+b_{i,j+1,k}
+b_{i,j-1,k}+b_{i,j,k+1}+b_{i,j,k-1}+6b_{i,j,k}\right]\nonumber\\
&&-\frac{1}{3(n+1)}\Omega_{m}a^3\Big(n\frac{c_1}{c_2^2}\Big)^{\frac{1}{n+1}}e^{-\frac{u_{i,j,k}}{n+1}}.
\end{eqnarray}
\end{widetext}
In principle, if we start from some high redshift, then the initial guess of $u_{i,j,k}$ could be chosen as the background value because we expect that any perturbations should be small then. For subsequent time steps we can use either the solution at the last time step or some analytical approximated solution as the initial guess (for Chameleon simulations we find the latter more convenient while for non-Chameleon simulations it is just the opposite).

\subsection{Non-Chameleon simulations}

The discretisation of the equation Eq.~(\ref{eq:sf_codeunit_nc}) for the non-Chameleon simulations is much easier, and similarly to the above we have
\begin{eqnarray}
L^{h}\left(v_{i,j,k}\right) &=& 0,
\end{eqnarray}
in which
\begin{widetext}
\begin{eqnarray}
L^{h}\left(v_{i,j,k}\right) &=&
\frac{1}{h^{2}}\frac{ac^2}{(BH_0)^2}\left[v_{i+1,j,k}+v_{i-1,j,k}+v_{i,j+1,k}
+v_{i,j-1,k}+v_{i,j,k+1}+v_{i,j,k-1}-6v_{i,j,k}\right]
\nonumber\\
&&+\Omega_{m}(\rho_{c,i,j,k}-1)-\frac{1}{3n(n+1)}\frac{c_2^2}{c_1}
\Omega_{m}a^3\left[3\left(\frac{1}{a^3}+4\frac{\Omega_\Lambda}{\Omega_{m}}\right)\right]^{n+2}v_{i,j,k},
\end{eqnarray}
\end{widetext}
with $v\equiv\dfR$, and
\begin{widetext}
\begin{eqnarray}
\frac{\partial L^{h}(v_{i,j,k})}{\partial v_{i,j,k}} =&&
-\frac{6}{h^{2}}\frac{ac^2}{(BH_0)^2}-\frac{1}{3n(n+1)}\frac{c_2^2}{c_1}\Omega_{m}a^3\left[3\left(\frac{1}{a^3}+4\frac{\Omega_\Lambda}{\Omega_{m}}\right)\right]^{n+2}.
\end{eqnarray}
\end{widetext}

\end{document}